\documentclass{aa} 

\usepackage{graphicx}
\usepackage{txfonts}
\usepackage{xcolor}
\usepackage{amsmath}
\usepackage{amsfonts}
\usepackage{natbib}
\usepackage{longtable}
\usepackage{booktabs}
\usepackage[amsmath,thmmarks]{ntheorem}
\theoremseparator{.}
\theorembodyfont{\normalfont} 
\newtheorem{assumption}{Assumption}{\bfseries}{\itshape}
\usepackage[normalem]{ulem}

\begin{document}

   \title{Miec: A Bayesian hierarchical model for the analysis of nearby young open clusters}
   \titlerunning{Miec}

    \author{J. Olivares\inst{1}
          \and H. Bouy \inst{1}
          \and L.~M. Sarro\inst{2}
          \and E. Moraux \inst{3}
          \and A. Berihuete\inst{4}
          \and P.A.B. Galli\inst{1}
          \and N. Miret-Roig\inst{1}
}

   \institute{Laboratoire d'astrophysique de Bordeaux, Univ. Bordeaux, CNRS, B18N, allée Geoffroy Saint-Hilaire, 33615 Pessac, France.
         \email{javier.olivares-romero@u-bordeaux.fr}
         \and
         Depto. de Inteligencia Artificial , UNED, Juan del Rosal, 16, 28040 Madrid, Spain 
         \and
         Univ. Grenoble Alpes, CNRS, IPAG, 38000 Grenoble, France
         \and
                Depto. Estad\'istica e Investigaci\'on Operativa. Universidad de C\'adiz, Avda. Rep\'ublica Saharaui s/n, 11510 Puerto Real, C\'adiz, Spain
        }

   \date{Received ; accepted }

 
  \abstract
   {The analysis of luminosity and mass distributions of young stellar clusters is essential to understanding the star-formation process. However, the gas and dust left over by this process extinct the light of the newborn stars and can severely bias both the census of cluster members and its luminosity distribution.}
   {We aim to develop a Bayesian methodology to infer, with minimal biases due to photometric extinction, the candidate members and magnitude distributions of embedded young stellar clusters.}
   {We improve a previously published methodology and extend its application to embedded stellar clusters. We validate the method using synthetically extincted data sets of the Pleiades cluster with varying degrees of extinction.}
   {Our methodology can recover members from data sets extincted up to $A_v\sim6$ mag with accuracies, true positive, and contamination rates that are better than 99\%, 80\%, and 9\%, respectively. Missing values hamper our methodology by introducing contaminants and artifacts into the magnitude distributions. Nonetheless, these artifacts vanish through the use of informative priors in the distribution of the proper motions.}
   {The methodology presented here recovers, with minimal biases, the members and distributions of embedded stellar clusters from data sets with a high percentage of sources with missing values (>96\%).}  


   \maketitle
%

\section{Introduction}
\label{section:introduction}

Stellar clusters are benchmarks against which the predictions of current theories of star formation and evolution can be compared and validated. In these comparisons, the youngest clusters play an important role because they still hold the imprints of the initial conditions of the molecular cloud from which they were formed. Unfortunately, because of the low star formation efficiency \citep[between 5\% and 30\%,][]{2007ARA&A..45..565M,2020arXiv200513401F}, most of the gas and dust of the parent molecular cloud remains in the vicinity of the newborn stars and extincts their light. This extinction can bias the census of candidate members and the population parameters derived from them, particularly the luminosity and mass distributions. For this reason, the propagation of the observational uncertainties and the characterization of methodological biases are unavoidable steps in the comparison of model predictions with observations.

In the last decade, diverse methodologies have been devised to determine the stellar census and properties of star-forming regions and open clusters \citep[e.g.,][]{2014A&A...563A..45S,2014A&A...561A..57K,2014ApJ...783..121G}. Following the second data release of the \textit{Gaia} mission  \citep{2018A&A...616A...1G}, hundreds of authors have used these high-quality measurements, in particular the highly discriminant parallax, to identify members in stellar clusters by applying diverse machine-learning techniques. The most popular techniques  are Gaussian mixture models, random forest classifiers \citep{Breiman2001}, DBSCAN \citep{Ester1996}, HDBSCAN \citep{10.1007/978-3-642-37456-2_14}, and density contrast using kernel density estimates. For example, \citet{2018A&A...618A..93C} used a modified version of UPMASK \citep{2014A&A...561A..57K} to identify members in 1229 clusters and \citet{2019AJ....158..122K} used HDBSCAN to identify 1900 clusters and comoving groups within 1 kpc. Other popular algorithms include \textit{Clusterix} \citep{2020MNRAS.492.5811B} and \textit{ASteCA} \citep{2015A&A...576A...6P}. While the former is a fully nonparametric method that determines cluster membership probabilities based on proper motions, the latter is a fully automated software that obtains cluster parameters like center coordinates and radius, together with luminosity functions and membership probabilities. To the best of our knowledge, \textit{ASteCA} is the only membership methodology from the literature that can deal with extinction, though based on theoretical isochrones that are known to face difficulties in reproducing the observed cluster photometric sequences in the low-mass domain (see the discussion in Sections 5.2 of \citealt{2015A&A...577A.148B} and \citealt{2019A&A...631A..57M}).

Although the censuses of stellar clusters have seen tremendous improvements thanks to the \textit{Gaia} data, the determination of population parameters based on these censuses is far from straight-forward because several biases can appear \citep{2018A&A...616A...9L}. Following the recommendations provided by the latter authors (see their Sect. 4.3), we improve the methodology of \citet[hereafter Paper I]{2018A&A...617A..15O}. The new code that we present here, which we call \textit{Miec}\footnote{The Pleiades were known as Miec by the Aztecs \citep{2002ASPC..282....3G}.}, is designed to simultaneously derive the census of members and the astrometric and photometric population distributions of embedded young stellar clusters.

The remainder of this paper is organized as follows. In Sect. \ref{section:methodology}, we review the methodology of Paper I and introduce the improvements, in particular those concerning the treatment of extinction. Section \ref{section:data} presents the synthetically extincted Pleiades clusters that is used in Sect. \ref{section:validation} to validate our methodology and analyze its biases. Finally, in Sect. \ref{section:conclusions}, we present our conclusions and perspectives.

\section{Methodology}
\label{section:methodology}

In this section, we describe the treatment of the photometric extinction as an improvement to the methodology of Paper I. We start by enumerating our assumptions and making a summary of the original model. Afterward, we describe the details of the two major improvements we present here: the treatment of extinction and a simplified model for equal-mass binaries. Also, since the publication of Paper I the code has been subject to a series of minor methodological and computational improvements that are described in Appendix \ref{appendix:improvements}.

\begin{assumption}
\label{assumption:gaussian}
Observed values are independent across stars and Gaussian distributed. The input catalog provides the necessary information to reconstruct these Gaussian distributions.
\end{assumption}

\begin{assumption}
\label{assumption:group}
The stellar cluster members share a common origin and therefore have similar properties but with an intrinsic dispersion. On the contrary, the field population has a heterogeneous origin, and therefore the cluster members can be probabilistically disentangled from it through statistical models constructed on features (i.e., observed values) of the astrometric and photometric spaces. The classification quality depends on the degree of overlapping between cluster and field populations and on the information provided by the features used in the models.
\end{assumption}

\begin{assumption}
\label{assumption:static_field}
Given that the field population overwhelmingly dominates the input catalog (typically >98\% of it), we assume that the field model inferred from the initial list of field sources can remain fixed during the inference of the cluster model.
\end{assumption}

\begin{assumption}
\label{assumption:independence}
The photometric and astrometric observed values of a source are independent of each other. Although incorrect, this assumption results in a moderate complexity model, with $n=66$ parameters. If we were to include the correlations between the astrometric and photometric parameters, the resulting model would be computationally intractable, with $\sim 500$ free parameters. Thus, this assumption is a reasonable compromise between the model's complexity and the computational time required to infer it. As a corollary, we assume that the photometric extinction does not affect the astrometric observables.   
\end{assumption}

\begin{assumption}
\label{assumption:binaries}
The stellar cluster is composed of only single stars and equal-mass binaries (hereafter EMBs). This assumption simplifies the treatment of the full spectra of binary mass-ratios.
\end{assumption}

\begin{assumption}
\label{assumption:extinction_map}
The input extinction map provides the upper limit, $A_{v,max}$, to the true extinction value, $A_v$, of each source, (i.e., $A_v \in [0,A_{v,max}]$). 
\end{assumption}

\begin{assumption}
\label{assumption:law}
The extinction follows the law of \citet{1989ApJ...345..245C}. We assume a value of $R_v=3.1$, which corresponds to the diffuse interstellar medium. This assumption neglects the effects of infrared excess, such as that  due to the presence of protoplanetary disks for example.  
\end{assumption}

\subsection{Summary of the original model}
\label{subsection:summary}

The aim of Paper I was to develop, test, and characterize a Bayesian hierarchical model designed to simultaneously identify members of nearby young open clusters and infer their population distributions (i.e., proper motions, and color and magnitude distributions). This model proceeds as follows.

The likelihood of a source is a mixture model with two components: the cluster likelihood, $\mathcal{L}_c$, and the field likelihood, $\mathcal{L}_f$; see Assumption \ref{assumption:group}. The weight or amplitude of the field component is parametrized as $\pi$. Because there are only two components, the cluster likelihood has amplitude 1-$\pi$. Each of these components has its own set of parameters, represented by $\boldsymbol{\theta_c}$ for the cluster, and $\boldsymbol{\theta_f}$ for the field. These parameters together with $\pi$, make the full set of model parameters: $\boldsymbol{\Theta}=\{\pi,\boldsymbol{\theta_c,\theta_f}\}$. 

We define the $N$-sources data set as: $\boldsymbol{\mathcal{D}}=\{d_i\}_{i=1}^N=\{\hat{\mu}_i,\hat{\Sigma}_i\}_{i=1}^N$, with $d_i$ the set of observables of the $i$-th source, which comprises the mean, $\hat{\mu}_i$, and the covariance matrix $\hat{\Sigma}_i$ of the observed quantities. Under Assumption \ref{assumption:gaussian}, the uncertainties are Gaussian and therefore the full likelihood can be written as

\begin{align}
    \mathcal{L}(\boldsymbol{\mathcal{D}}\mid\boldsymbol{\Theta})&\equiv\prod_{i=1}^N \mathcal{L}(d_i\mid\boldsymbol{\Theta})\nonumber \\
    &=\prod_{i=1}^N\pi\cdot\mathcal{L}_f(d_i\mid\boldsymbol{\theta}_f)+(1-\pi)\cdot\mathcal{L}_c(d_i \mid\boldsymbol{\theta}_c),
\end{align}
which is equivalent to Eq. 3 of Paper I. 

Assumption \ref{assumption:independence} allows us to factorize the likelihood of the $i$-th source into the astrometric and photometric parts as

\begin{align}
\label{eq:likelihood}
    \mathcal{L}(d_i\mid\boldsymbol{\Theta})&= \pi\cdot\mathcal{L}_f^A(d_i\mid\boldsymbol{\theta}_f^A)\cdot\mathcal{L}_f^P(d_i\mid\boldsymbol{\theta}_f^P) \nonumber \\
    &+(1-\pi)\cdot\mathcal{L}_c^A(d_i\mid\boldsymbol{\theta}_c^A) \cdot \mathcal{L}_c^P(d_i\mid\boldsymbol{\theta}_c^P),
\end{align}
where the superscripts $A$ and $P$ stand for astrometry and photometry, respectively.

The cluster photometry is modeled as a mixture of two components (see Assumption \ref{assumption:binaries}): single-stars, denoted with subscript $s$, and EMB, denoted with subscript $b$. This mixture's weights are given by $\pi_s$ for the single-star population and 1-$\pi_s$ for the EMB population. With these definitions, the likelihood of the $i$-th source can be written as

\begin{align}
    \mathcal{L}(d_i\mid\boldsymbol{\Theta})&=\pi\cdot\mathcal{L}_f^A(d_i\mid\boldsymbol{\theta}_f^A)\cdot\mathcal{L}_f^P(d_i\mid\boldsymbol{\theta}_f^P)
    +(1-\pi)\cdot\mathcal{L}_c^A(d_i\mid\boldsymbol{\theta}_c^A)\cdot\nonumber \\
    &\left[\pi_s\mathcal{L}_{c,s}^P(d_i\mid\boldsymbol{\theta}_{c,s}^P)+ (1-\pi_s)\mathcal{L}_{c,b}^P(d_i\mid\boldsymbol{\theta}_{c,b}^P)\right].
\end{align}

The photometric likelihoods of both single and EMB populations are parametrized with the source true color index, $CI$. However, as the latter is a source-level parameter we marginalize it with the aid of a population-level prior, $p(CI\mid \phi)$. The prior parameters, $\phi$, are included in the set of cluster parameters (i.e., $\phi\in \boldsymbol{\theta}_c$). The marginalization of the $CI$ parameter is done as follows,

\begin{align}
\label{eq:marg_lik}
    \mathcal{L}(d_i\mid\boldsymbol{\Theta})&=\int\mathcal{L}(d_i\mid CI,\boldsymbol{\Theta})\cdot p(CI\mid \phi)\cdot {\rm d}CI \nonumber \\
    &=\pi\cdot\mathcal{L}_f^A(d_i\mid\boldsymbol{\theta}_f^A)\cdot\mathcal{L}_f^P(d_i\mid\boldsymbol{\theta}_f^P) +(1-\pi)\cdot\mathcal{L}_c^A(d_i\mid\boldsymbol{\theta}_c^A) \cdot\nonumber \\
    &\left[\pi_s \int\mathcal{L}_{c,s}^P(d_i\mid CI,\boldsymbol{\theta}_{c,s}^P)\cdot p(CI\mid \phi)\cdot {\rm d}CI\right. \nonumber \\
    &\left.+(1-\pi_s)\cdot \int \mathcal{L}_{c,b}^P(d_i\mid CI,\boldsymbol{\theta}_{c,b}^P)\cdot p(CI\mid \phi)\cdot {\rm d}CI
    \right].
\end{align}

The explicit definitions of all terms in the previous equation can be found in Paper I, together with the prior distribution for their parameters.

\subsection{Treatment of extinction}
\label{subsection:extinction}
Our treatment of the photometric extinction follows a similar approach to that taken in Paper I for the marginalization of the true color index of each source. We include the true extinction, $A_v$, of each source as a model parameter that we marginalize with the aid of a uniform prior, $p(A_v\mid\psi)$, with $\psi$ its parameters, the support of which is provided by the extinction map.

Thus, in the extinction module, the likelihood of the $i$-th source given the original model parameters, $\boldsymbol{\Theta}$, and the new extinction parameters, $\psi$, is

\begin{align}
\label{eq:extinction}
    \mathcal{L}(d_i\mid\boldsymbol{\Theta},\psi)&\equiv\int\mathcal{L}(d_i,A_v\mid\boldsymbol{\Theta})\cdot {\rm d}A_v \nonumber \\
    &=\int\mathcal{L}(d_i\mid A_v,\boldsymbol{\Theta})\cdot p(A_v\mid \psi)\cdot {\rm d}A_v \nonumber \\
    &\propto\int_{0}^{A_{v,max}}\mathcal{L}(d_i\mid A_v,\boldsymbol{\Theta})\cdot {\rm d}A_v,
\end{align} 
\noindent
where the term $\mathcal{L}(d_i\mid A_v,\boldsymbol{\Theta})$ is given by an expression similar to that of Eq. \ref{eq:marg_lik}. However, in the latter, the true photometry is reddened according to the assumed extinction law (see Assumption \ref{assumption:law}) using the values of $A_v$ in the support of the prior $p(A_v\mid\psi)$. These integrals, one for each source, are computed numerically as in Paper I.

We note that the marginalization of the true extinction of the source  differs from that of its true color index in the sense that the prior parameters $\psi$ (i.e., the lower and upper bounds of the uniform prior) are not inferred by the model but are provided by the extinction map (see Assumption \ref{assumption:extinction_map}). Although using these limits rather than inferring them diminishes the complexity of the model, the marginalization integrals severely increase the computation time. The latter grows linearly with the number of evaluation steps of the integral, with roughly every step taking the same amount of time as one likelihood evaluation of our basic nonextinction module. We heuristically set the grid steps to 20, which proved to be a good compromise between the accuracy of the integral and the total computing time.

\subsection{Simplified model for equal-mass binaries}
\label{subsection:simplified_emb}
Stellar clusters are known to host a non-negligible fraction of stars coupled in binaries or multiple systems. Observationally, these multiple systems are seen as a spread of the cluster photometric sequence in color--magnitude diagrams. In particular, EMBs are located in a usually sharper sequence 0.75 magnitudes brighter than the single star sequence. In Paper I, the EMBs were modeled with their own parallel photometric sequence and astrometric Gaussian mixture model (GMM). The new EMB module allows us to model the EMBs with a parallel photometric sequence, but it forces the astrometric GMM to be the same as that of the single stars. This simplification reduces the number of astrometric parameters by half and therefore also the model complexity, which helps to decrease the computation time of an expensive model like the extinction one. Although the simplification introduced by this module implies discarding $\sim2$\% of the candidate members recovered with the original EMB module, it nonetheless recovers the same fraction of EMBs in the cluster population (for more details see Appendix \ref{appendix:benchmark_solution}).

\section{Data set}
\label{section:data}

We validate our methodological and computational improvements on the Pleiades data set used in Paper I. This data set corresponds to the $10^5$ most probable candidate members of the Pleiades DANCe catalog \citep{2015A&A...577A.148B}. We use the same representation space as in Paper I, which consists of the proper motions, the color index $i-K$, and the photometric bands $Y, J, H$, and $K$. Table \ref{table:catalog} gives a summary of this data set, which we hereafter refer to as the original one.

As described in Paper I, the color index in the representation space must be carefully chosen to minimize the number of missing values, maximize the wavelength interval, and fulfill the injective requirement in the splines that model the photometric magnitudes as functions of the color index. The previous requirements prevent us from using, for example, the colors $Y-J$ or $J-K$, which result in fewer missing values than the $i-K$ color, but have narrower wavelength intervals and produce vertical photometric sequences that cannot be modeled with injective functions.

To validate the new modules of our methodology (see Sects. \ref{subsection:extinction} and \ref{subsection:simplified_emb}) we generate synthetically extincted Pleiades data sets by reddening the original photometry with synthetic extinction maps (more below). Each of the latter provides the upper limit of the source extinction, $A_{v,max}$ (see Assumption \ref{assumption:extinction_map}). If the source is a candidate member (according to the results of Paper I), then its true extinction is uniformly sampled between zero and the upper limit provided by the map. Given the proximity of the Pleiades cluster \citep[$\varpi=7.44 \pm 0.08$ mas,][]{2017A&A...598A..48G}, the population of foreground contaminants is negligible ($\lesssim$0.2\% according to the distribution of \textit{Gaia} DR2 parallaxes for sources in the same sky region of our catalog). Thus, if the source is classified as a field contaminant according to Paper I, then we use the upper limit provided by the map as its true extinction. Finally, the photometry is reddened using the true $A_v$ of each source, together with the extinction law of \citet{1989ApJ...345..245C} with $R_v=3.1$, and the effective wavelength of each photometric passband (i.e., $i,Y,J,H,K=\{$7630\AA, 10310\AA, 12500\AA, 16500\AA, 21500\AA$\}$).

\begin{table}[ht!]
\caption{Properties of the original data set.}
\label{table:catalog}
\centering
\begin{tabular}{c|c|c|c}
\toprule
{} &   Min. &  Max. &  Missing [\%] \\
Observables &        &       &              \\
\midrule
pmra $\rm{[mas \cdot yr^{-1}]}$  & -100.0 &  99.9 &          0.0 \\
pmdec $\rm{[mas \cdot yr^{-1}]}$      & -100.0 & 100.0 &          0.0 \\
i-K [mag]        &    0.9 &   8.0 &         96.4 \\
Y   [mag]        &    8.3 &  22.2 &         22.2 \\
J   [mag]        &    4.0 &  20.6 &          7.2 \\
H   [mag]        &    3.0 &  20.3 &          7.5 \\
K   [mag]        &    2.6 &  21.0 &          4.9 \\
\bottomrule
\end{tabular}

\end{table}

\subsection*{Synthetic extinction maps}
We need synthetic extinction maps that produce $A_{v,max}$ probability distributions with the following properties. First, they must resemble those of real cases without being so specific that the validation loses generality. Second, they must allow us to control the number of sources within a given interval of extinction, which is needed to ensure good-number statistics over the probed extinction values. We restrain ourselves from using the extinction maps of real regions because in that case, our validation would lose generality because the resulting $A_{v,max}$ probability distributions would be highly dependent on the spatial distributions and extinction maps of those particular regions.

We decided to generate synthetic maps using GMMs on the space of synthetic sky coordinates. On the one hand, the GMMs have enough flexibility \citep[see][for a review of the use of mixture models in astronomy]{2017arXiv171111101K} to mimic the possibly cloudy structures present in nearby star-forming regions, but on the other, the synthetic sky coordinates in combination with the parameters of the GMM allow us to control the fraction of sources above a certain value of extinction (i.e., the shape of the $A_{v,max}$ cumulative distribution). We note that our aim is to produce realistic and generic $A_{v,max}$ probability distributions rather than realistic 2D extinction maps. Furthermore, as the sky coordinates are not part of the representation space of our methodology, they do not influence our results beyond setting the $A_{v,max}$ of the sources and allowing us to control their numbers within a given extinction interval. Therefore, we define the $A_{v,max}$ of our extinction maps as 

\begin{equation}
\label{equation:gmm_map}
    A_{v,max}(\alpha,\delta)=\sum_{i=1}^k \frac{w_i}{C_{\{\mu_i,\Sigma_i\}}}\cdot\mathcal{N}(\alpha,\delta \mid \mu_i,\Sigma_i),
\end{equation}

where $\alpha$ and $\delta$ are the synthetic sky coordinates, $k$ is the number of Gaussian components, and $\boldsymbol{w}$, $\boldsymbol{\mu}$, and  $\boldsymbol{\Sigma}$ are the weights, means, and diagonal covariance matrices of the GMM, respectively. The factor $C_{\{\mu_i,\Sigma_i\}}$ is a constant that is equal to the density of the Gaussian distribution at its mean; it allows us to set $w_i$ as the maximal extinction for the $i$th Gaussian component. The synthetic sky coordinates of the sources follow a bivariate uniform distribution, with $\alpha,\delta$ in the interval [-10\degr,10\degr] in R.A. and Dec. 

\begin{figure}[ht!]
    \centering
    \includegraphics[width=\columnwidth,page=1]{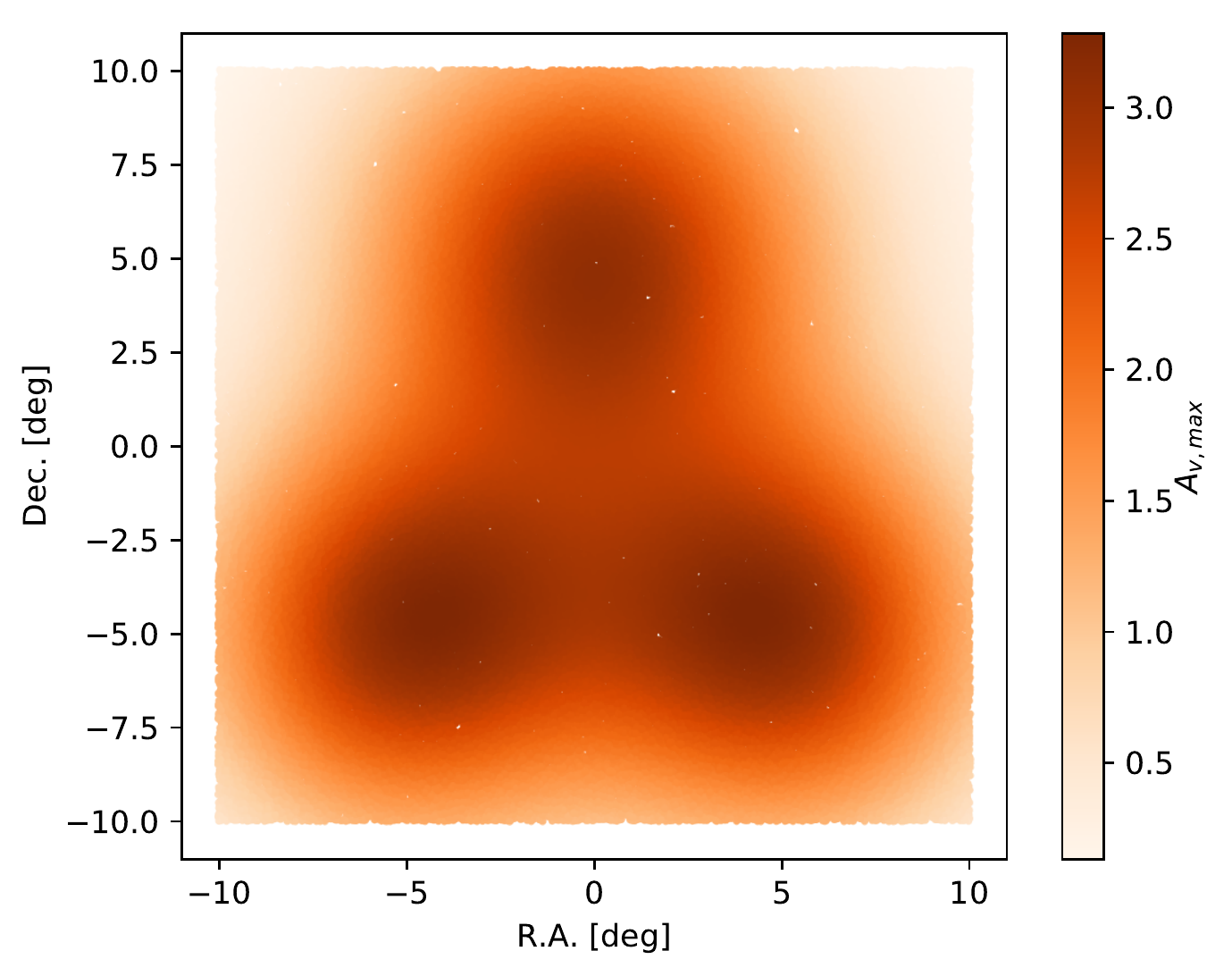}
     \includegraphics[width=\columnwidth,page=2]{Figures/Extinction_map_1.pdf}
     \caption{Top panel: Synthetic 2D extinction map, MAP$_1$. Bottom panel: Color--magnitude diagram showing the original photometry of the $10^5$ most probable members from the Pleiades DANCe catalog, the reddened one, and the extinction vectors as obtained from the map shown in the upper panel.}
\label{fig:extinction_map_1}
\end{figure}

After testing several configurations of parameter values for  Eq. \ref{equation:gmm_map}, we found that GMMs with three components and the following parameter values produce what we consider are representative case scenarios for the application of our methodology (more below).

\begin{align}
\rm{MAP}_0:& \boldsymbol{w}=\{3,2,1\},\ \ \boldsymbol{\mu}=\{[5,5],[5,-5],[-5,-5]\}, \nonumber \\
                   & \boldsymbol{\Sigma}=\{[3,3],[2,2],[1,1]\},\nonumber \\
\rm{MAP}_1:& \boldsymbol{w}=\{3,3,3\},\ \ \boldsymbol{\mu}=\{[0,5],[5,-5],[-5,-5]\}, \nonumber \\
                   & \boldsymbol{\Sigma}=\{[20,20],[15,15],[15,15]\},\nonumber \\
\rm{MAP}_2:& \boldsymbol{w}=\{6,6,6\},\ \ \boldsymbol{\mu}=\{[5,5],[5,-5],[-5,-5]\}, \nonumber \\
           & \boldsymbol{\Sigma}=\{[12,12],[8,8],[4,4]\}. \nonumber
\end{align}

With the previous maps, we reddened the original data set and obtained three synthetically extincted data sets. In the following, we refer to these data sets as MAP$_0$, MAP$_1$, and MAP$_2$ according to the extinction map used to redden each of them.

\begin{figure}[ht!]
    \centering
    \includegraphics[width=\columnwidth]{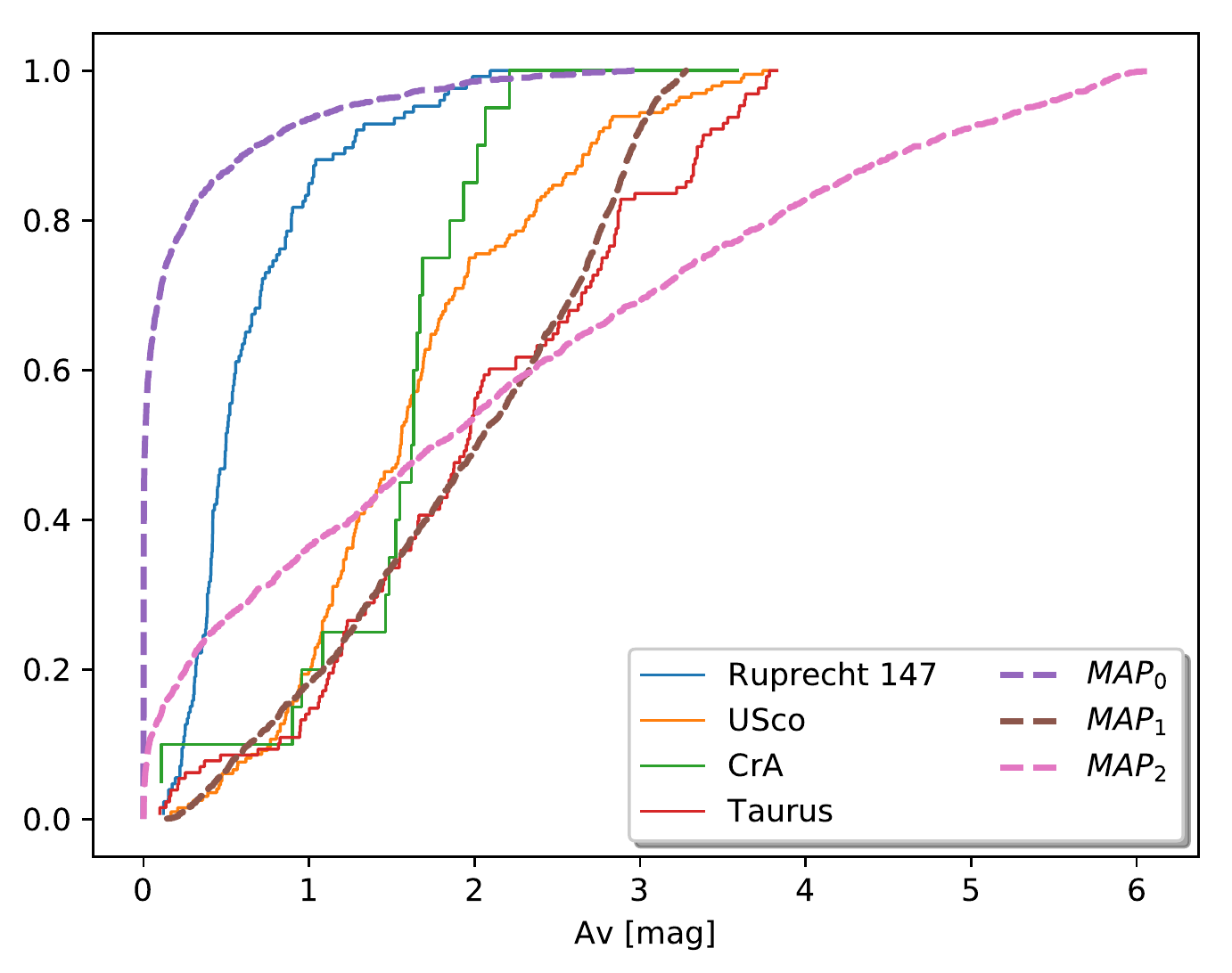}
     \caption{Cumulative distribution of $A_v$ for the members of the open cluster Ruprecht 147, and the star-forming regions of Corona Australis, Upper Scorpius, and Taurus. The $A_{v,max}$ cumulative distributions yielded by our synthetic extinction maps for the Pleiades members found in Paper I are also shown with dashed lines.}
\label{fig:cumulatives}
\end{figure}

As an example, Fig. \ref{fig:extinction_map_1} shows the extinction MAP$_1$ as well as the original, and reddened photometry. In Fig. \ref{fig:cumulatives} we show the $A_v$ cumulative distributions of the open cluster Ruprecht 147 \citep{2019A&A...625A.115O}, the star-forming regions of Corona-Australis \citep{2020A&A...634A..98G}, Upper Scorpius (Miret-Roig et al. in preparation), and Taurus (Olivares et al. in preparation), as well as those of the $A_{v,max}$ obtained by our synthetic extinction maps. The $A_v$ values of the members in the previous star-forming regions and stellar clusters were obtained by transforming their \textit{Gaia} DR2 \texttt{a\_g\_val} using the relation $A_G/A_v=0.789\pm0.005$ \citep{2019ApJ...877..116W}.

As can be observed in Fig. \ref{fig:cumulatives}, our synthetic extinction maps reproduce the $A_{v,max}$ distribution of low-, intermediate-, and high-extinction regions. The extinction map $MAP_0$ mimics a low-extinction region, with extinction values similar to those expected in old open clusters that have expelled the majority of the remnant gas and dust. The extinction map $MAP_1$ mimics an intermediate-extinction region similar to those of Taurus and Upper Scorpius. We notice that the three Gaussian components of our synthetic extinction maps are enough to reproduce the overall trend of extinction present in the Taurus region. Although the use of more Gaussian components would certainly result in a closer match to the observed extinction distribution, as mentioned above, it would imply a loss of generality in the validation. Finally, the extinction map MAP$_2$, mimics a high-extinction region with values larger than those observed in Taurus or Upper Scorpius. We created this extinction map to validate our methodology beyond the extinction values of candidate members found with current membership methodologies that, although taking into account the photometry, are unable to deal with extinction \citep[see e.g.,][]{2014A&A...563A..45S,2014A&A...561A..57K,2019A&A...625A.115O}.

\section{Validation}
\label{section:validation}

We ran our new extinction and simplified EMB modules (hereafter extinction+EMB module; see Sect. \ref{section:methodology}) on the synthetically reddened MAP$_{0,1,2}$ data sets (see Sect. \ref{section:data}). Each of these runs takes $\sim$10 days on a computing server with 45 CPUs at 2.1 GHz and 8 GPUs NVIDIA GForce. We also ran our improved methodology on the original nonextincted data set using the simplified EMB module but without the extinction one (hereafter basic+EMB module). This latter solution is used as a benchmark against which the results of the extinction+EMB module are compared. The details of this benchmark solution and the validation of the EMB module are given in Appendix \ref{appendix:benchmark_solution}.

In the following, we first validate the capability of our methodology to recover the extincted members. Afterward, we assess the biases that extinction introduces in the inferred population parameters and the magnitude distributions in particular.

\subsection{Quality of the classifier}
\label{subsection:classifier}

We measure the quality of our classifier using the confusion matrices that result from the inferred membership probabilities and the optimum probability threshold for classification. As described in Paper I, our methodology returns, for each source in the data set and for each combination of model parameters $\Theta$, two membership probabilities: one of belonging to the cluster and another of being an EMB. Therefore, the MCMC sampling of the posterior distribution of the model parameters results in  distributions for the cluster and EMB membership probabilities. 

As in Paper I, here we classify a source as a candidate member if 86\% of its membership probability distribution is above the optimum probability threshold $p_t$ (i.e., if $P_{\mu} + P_{\sigma} > p_t$, where the first and second terms on the left-hand side are the mean and standard deviation of the distribution of membership probability).
The optimum probability threshold, $p_t$, corresponds to the probability threshold that maximizes the accuracy of the classifier (ACC, see Eq. 10 of Paper I) when applied over data where the true classes are known (i.e., cluster or field populations, single-stars or EMB). We notice that this classification threshold is not part of the model but is found a posteriori based on a specific criterion: the maximum classification accuracy. We find the optimum probability threshold, $p_t$, with three different strategies. 

The first strategy uses the learned cluster and field models to generate synthetic data where the true classes are known, then it runs the method over these data and computes the optimum probability threshold. The second and third strategies instead use the class labels of the original nonextincted data set. While the second strategy uses all sources independently of their extinction value, the third one splits the data into bins of one magnitude of extinction and computes an optimum probability threshold for each bin. Appendix \ref{appendix:classifier} describes the details of these three strategies and the results they produce when applied to the synthetic data sets MAP$_{0,1,2}$. The three strategies are similarly good at recovering the cluster members, but the third one results in cleaner samples due to a reduced contamination rate.

When used as a classifier, our extinction methodology recovers candidate members from datasets with extinction up to $A_{v,max}\sim6$ mag, which, under the information content of the present data set, represents an upper limit to our method. We observe that the true positive rate and contamination rate depend largely on the observed status of the color index and the maximum value of extinction. In the most extincted of our data sets (i.e., MAP$_2$), the contamination rate measured with our third strategy on the subset of sources with observed color index is $\lesssim$4\% and increases to $\sim$20\% for the subset of sources with a missing color index. The detailed analysis of classification thresholds by bins of extinction allowed us to improve the true positive rate and contamination rate to values that are better than 83\% and 9\%, respectively. Despite this success, our method faces two problems: an increased contamination rate due to sources with missing values and a reduced true positive rate in sources with high extinction values. In the next section, we analyze the impact that these missing-value contaminants have on the population distributions.

\subsection{Population distributions}
\label{subsection:parameters}
In this section, we analyze the ability of our extinction methodology to recover the parameters of the benchmark solution, particularly those of the magnitude distributions. As the population distributions are obtained directly from the inferred parameters of the model, they are independent of the probability thresholds and strategies to obtain them. We start by comparing the posterior distribution of the model parameters, the resulting proper motions, and color index distributions, and finally we analyze the accuracy of the inferred magnitude distributions.

In the models inferred from the MAP$_{0,1,2}$ data sets, between 15\% and 33\% of the parameters have a maximum-a-posteriori value that is discrepant, beyond $3\sigma$ (with $\sigma$ being the square root of the sum of the variances of the parameter posterior distributions), from the value of the benchmark solution. The most discrepant parameters are those associated with the intrinsic dispersion of the photometric sequence and with the fractions and covariance matrices of the proper motions GMM. The previous discrepancies have their origin in the contaminants introduced in the extinction model, as explained below.

In our methodology, all sources in the data set contribute to the cluster model proportionally to their membership probability. Consequently, the contaminants of the model, which are by definition the field population, contribute to the broadening of both the photometric sequence and the distribution of proper motions (i.e., the most discrepant parameters), and therefore to the shifting of their parameters. As described in the previous section and Appendix \ref{appendix:classifier}, the majority of the contaminants have their origin in a missing color index. Therefore, the broadening of the model and the shifting of its parameters results from the lack of constraining information produced by the high percentage of sources
with missing values (see Table \ref{table:catalog}). 

\begin{figure}[ht!]
    \centering
    \includegraphics[width=\columnwidth,page=1]{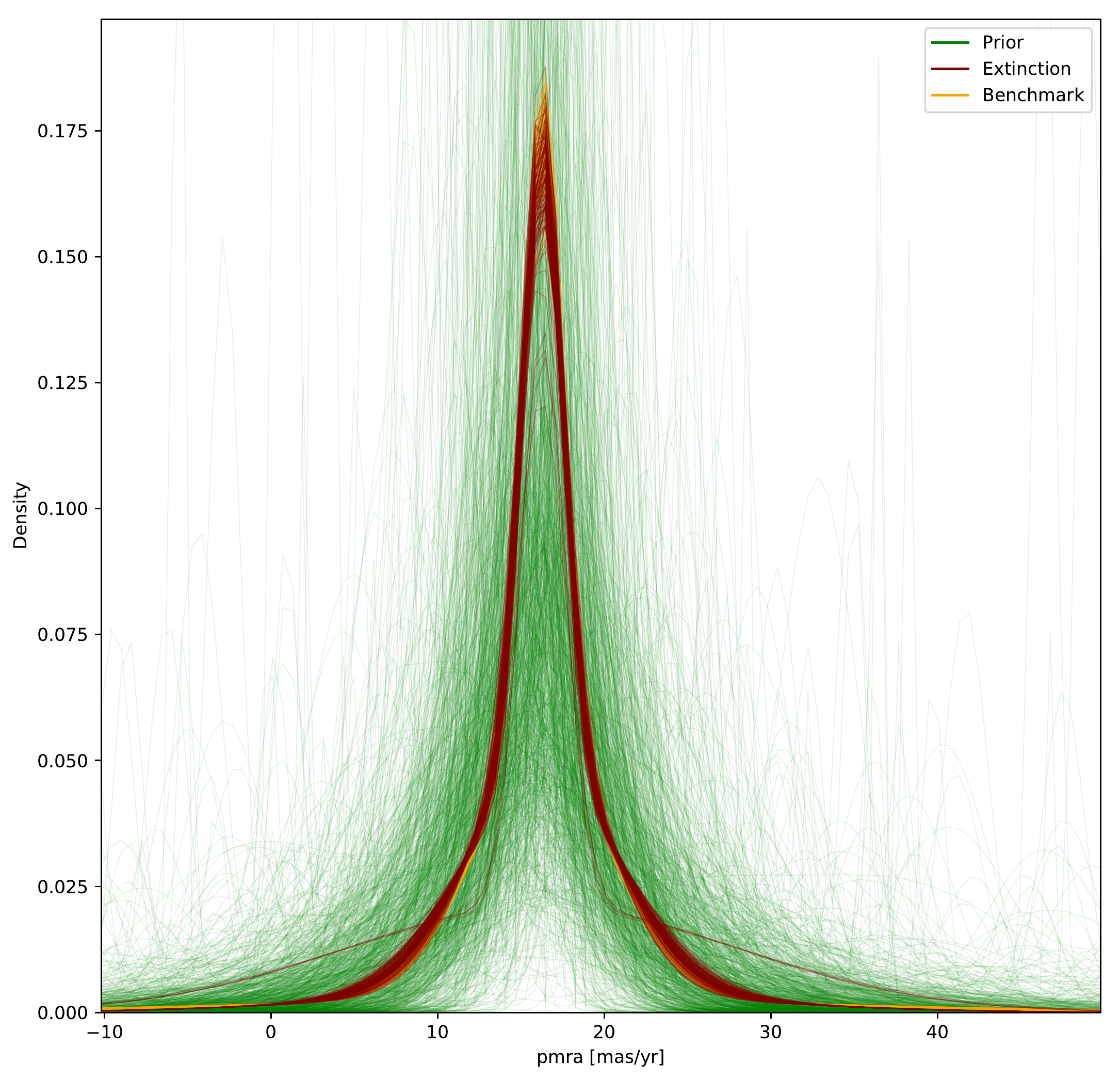}
    \includegraphics[width=\columnwidth,page=2]{Figures/Model_comparison_0.pdf}
     \caption{Comparison of the distributions of proper motions (pmra, upper panel) and color index (lower panel) inferred from the MAP$_0$ data set (maroon lines) and those of the benchmark solution (orange lines). The prior distribution is shown with green lines.}
\label{fig:comparison_ppm_clr_0}
\end{figure}

\begin{figure}[ht!]
    \centering
    \includegraphics[width=\columnwidth,page=1]{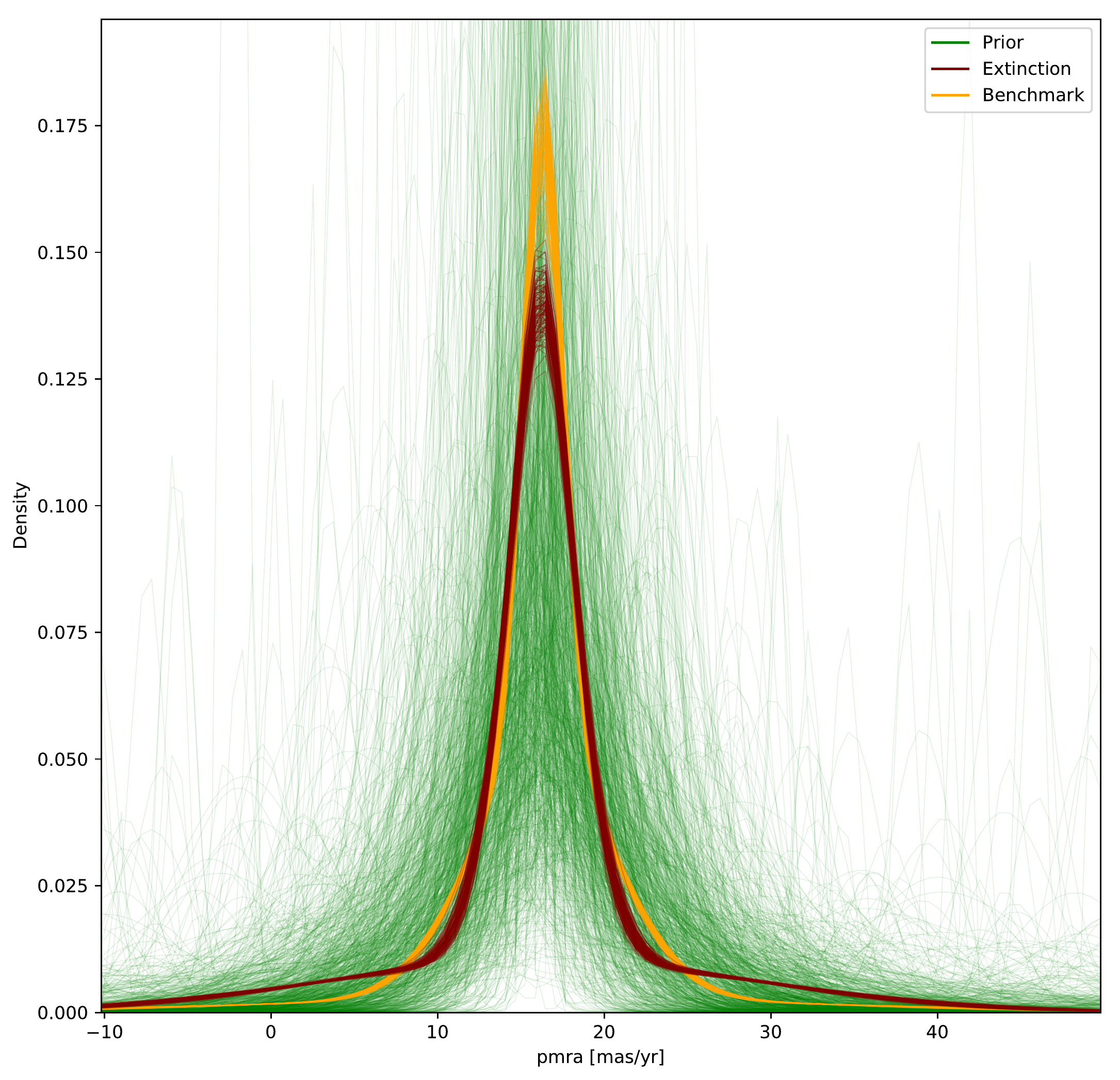}
    \includegraphics[width=\columnwidth,page=2]{Figures/Model_comparison_1.pdf}
     \caption{Same as Fig. \ref{fig:comparison_ppm_clr_0} but for the MAP$_1$ data set.}
\label{fig:comparison_ppm_clr_1}
\end{figure}

\begin{figure}[ht!]
    \centering
    \includegraphics[width=\columnwidth,page=1]{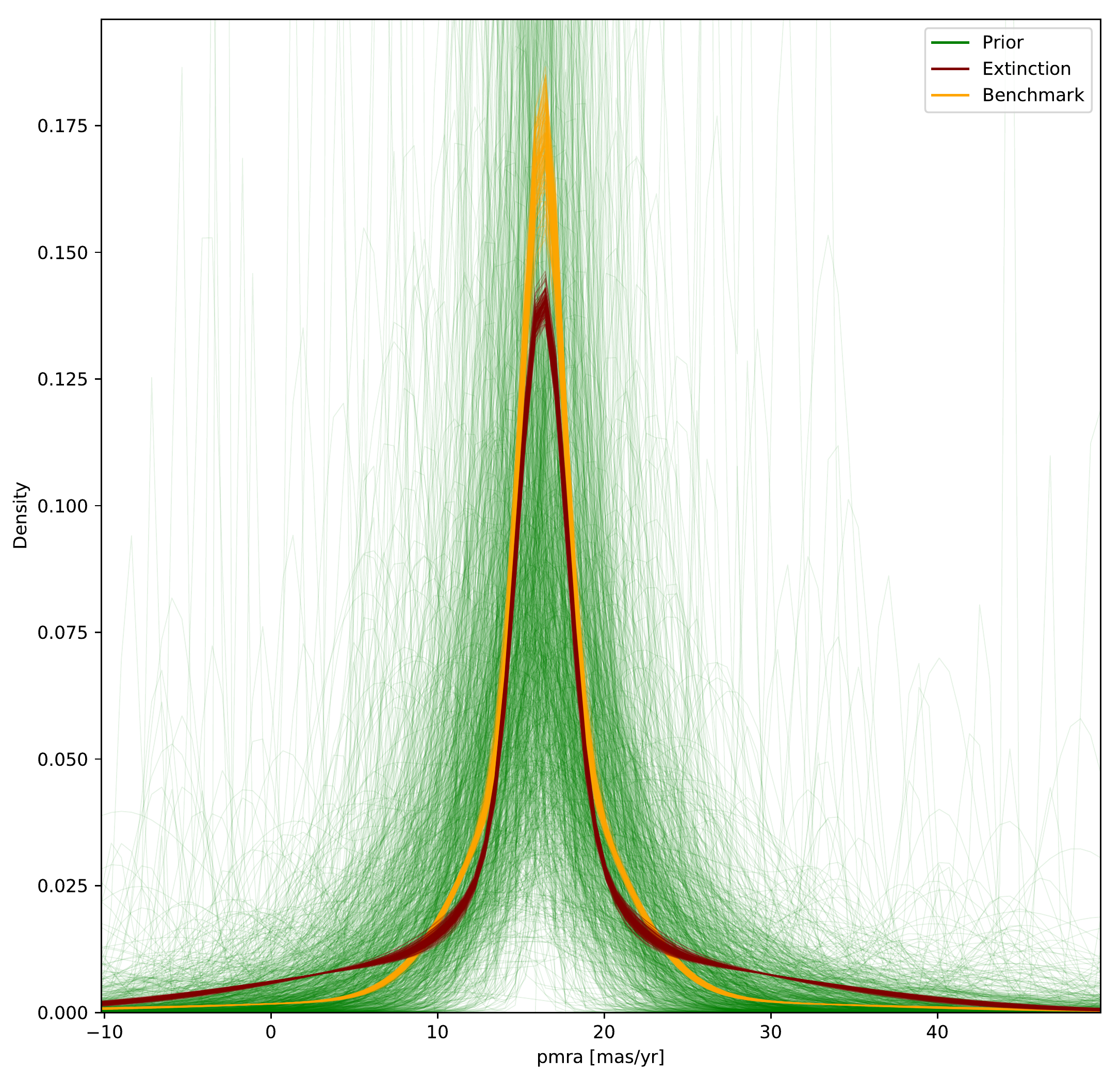}
    \includegraphics[width=\columnwidth,page=2]{Figures/Model_comparison_2.pdf}
     \caption{Same as Fig. \ref{fig:comparison_ppm_clr_0} but for the MAP$_2$ data set.}
\label{fig:comparison_ppm_clr_2}
\end{figure}

Figures \ref{fig:comparison_ppm_clr_0} to \ref{fig:comparison_ppm_clr_2} show the marginal posterior distributions of the proper motions and color index inferred from the MAP$_{0,1,2}$  data sets. These distributions are constructed from samples of the posterior distribution of the model parameters, and each line shows a single sample of the MCMC. For comparison purposes, the figures also display samples of the posterior distributions of the benchmark solution (orange lines) and the prior distribution (green lines). 

As can be observed from these figures, the model learned from the MAP$_0$ data set is almost indistinguishable from the benchmark solution in both proper motions and color index, despite having a 15\% of discrepant parameters, as discussed above. On the contrary, the models learned from the MAP$_{1,2}$ data sets show wider wings in the proper motions distributions. These wings result from the contaminants shown as the dispersed population of false positives in the upper panels of Figs. \ref{fig:members_map_1} and \ref{fig:members_map_2}. 

The lower panels of Figs. \ref{fig:comparison_ppm_clr_0} to \ref{fig:comparison_ppm_clr_2} show a gradual trend in the discrepancy between the color index distribution inferred from the MAP$_{0,1,2}$ data sets and that of the benchmark solution, with the discrepancy increasing with the extinction value. The features in the color distribution are gradually smoothed until, at the maximum extinction of $A_v\sim6$ mag (i.e., the MAP$_2$ data set), the features at i-K $\sim$ 1.5, 2, and 4.5 mag are completely smoothed out. The previous effects are a direct consequence of the increased number of contaminants with increasing extinction value. Despite these effects, the bulk of the proper motions and color index distributions are recovered without significant shifts, but only with a general broadening. 

The precision of the inferred population distributions, which is shown as the width of the posterior samples in Figs. \ref{fig:comparison_ppm_clr_0} to \ref{fig:comparison_ppm_clr_2}, remains similar to that of the benchmark solution. The only exception is the blue side of the color index distribution inferred from the MAP$_2$ data set, between 0.8 mag and 3 mag, where the lines are less jammed. This loss of precision is a direct consequence of the lack of information resulting from a large number of missing values in the Y band (see Table \ref{table:catalog}, and the black and blue lines in the upper panel of Fig. \ref{fig:benchmark_magnitude}) in combination with the high values of extinction.

The previous analysis indicates that the model inferred with our extinction methodology shows a broadening that increases with increasing values of extinction. In the MAP$_0$ data set, this broadening is negligible. In the MAP$_1$, it is only observed in the distribution of proper motions, and in the MAP$_2$ data set, it is observed in both the proper motions and the color index distribution.

\subsubsection*{Magnitude distributions}
\begin{figure}[ht!]
    \centering
    \resizebox{\columnwidth}{!}{
    \includegraphics[page=3]{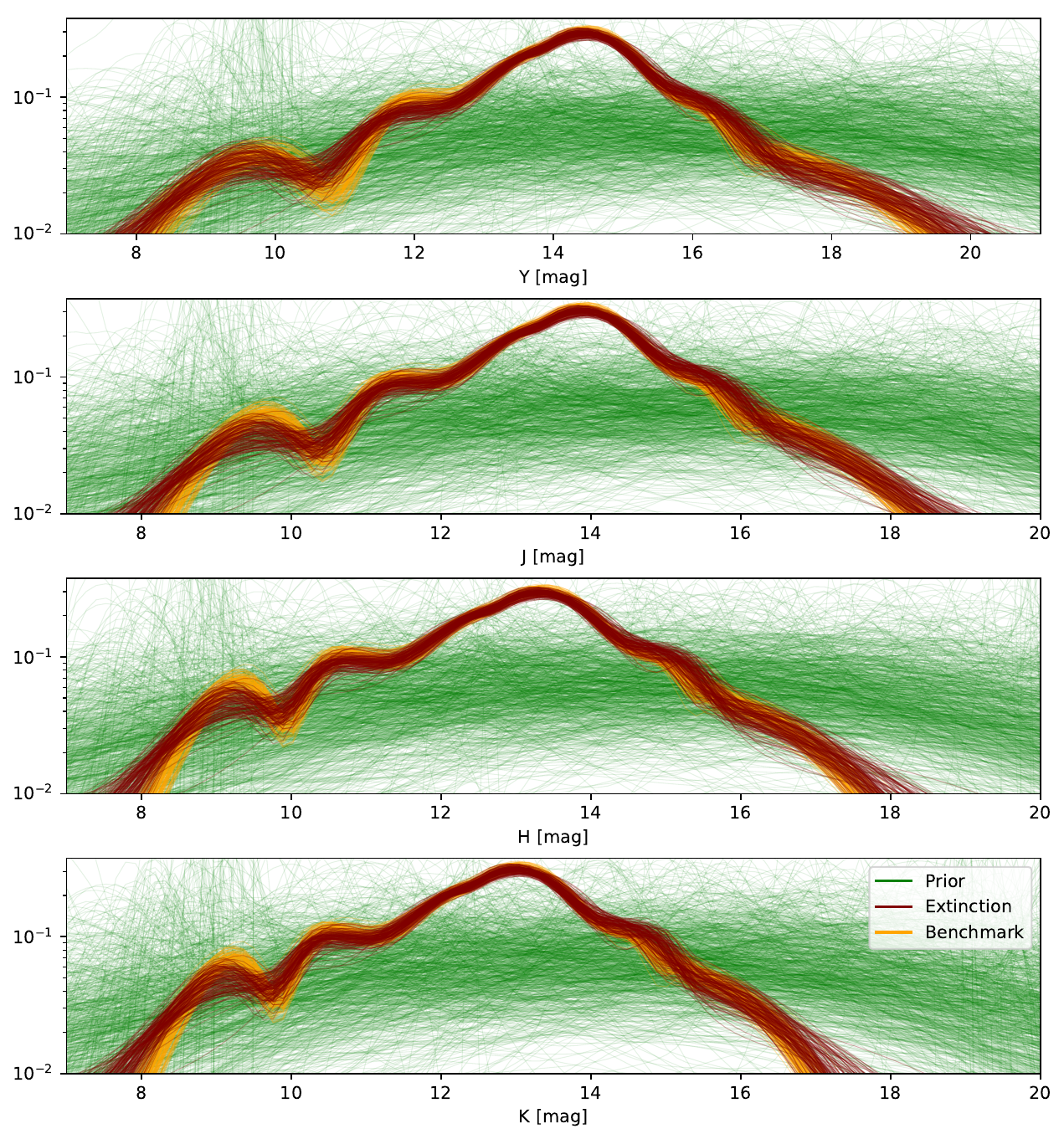}
    }
     \caption{Comparison of the Y, J, H, and K magnitude distributions learnt from the original (orange lines) and MAP$_0$ (maroon lines) data sets. The prior distribution is shown with green lines.}
\label{fig:comparison_YJHK_0}
\end{figure}

\begin{figure}[ht!]
    \centering
    \resizebox{\columnwidth}{!}{
    \includegraphics[page=3]{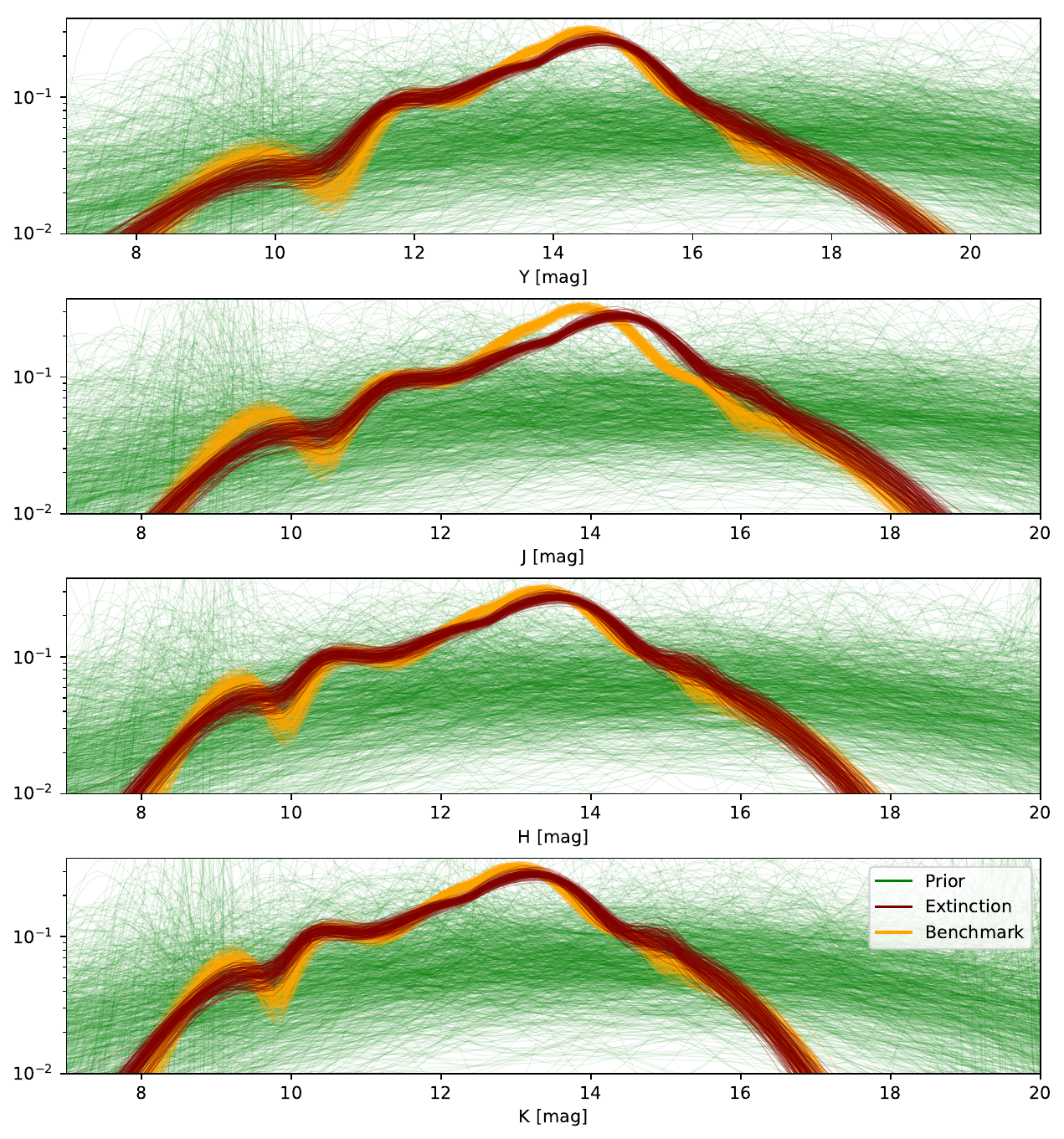}
    }
     \caption{Same as Fig. \ref{fig:comparison_YJHK_0} but for the MAP$_1$ data set.}
\label{fig:comparison_YJHK_1}
\end{figure}

\begin{figure}[ht!]
    \centering
    \resizebox{\columnwidth}{!}{
    \includegraphics[page=3]{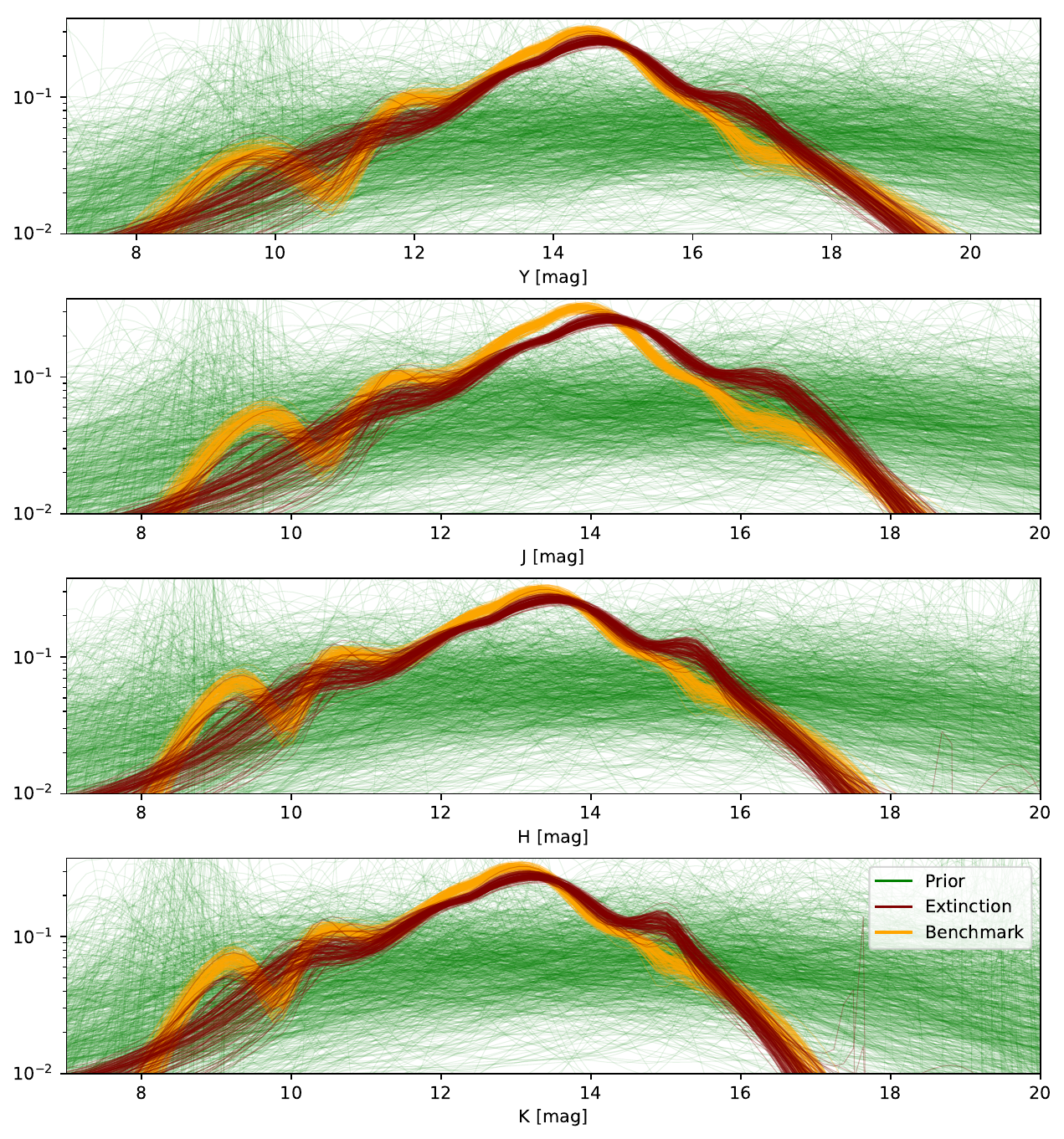}
    }
     \caption{Same as Fig. \ref{fig:comparison_YJHK_0} but for the MAP$_2$ data set.}
\label{fig:comparison_YJHK_2}
\end{figure}

\begin{figure}[ht!]
    \centering
    \resizebox{\columnwidth}{!}{
    \includegraphics[page=1]{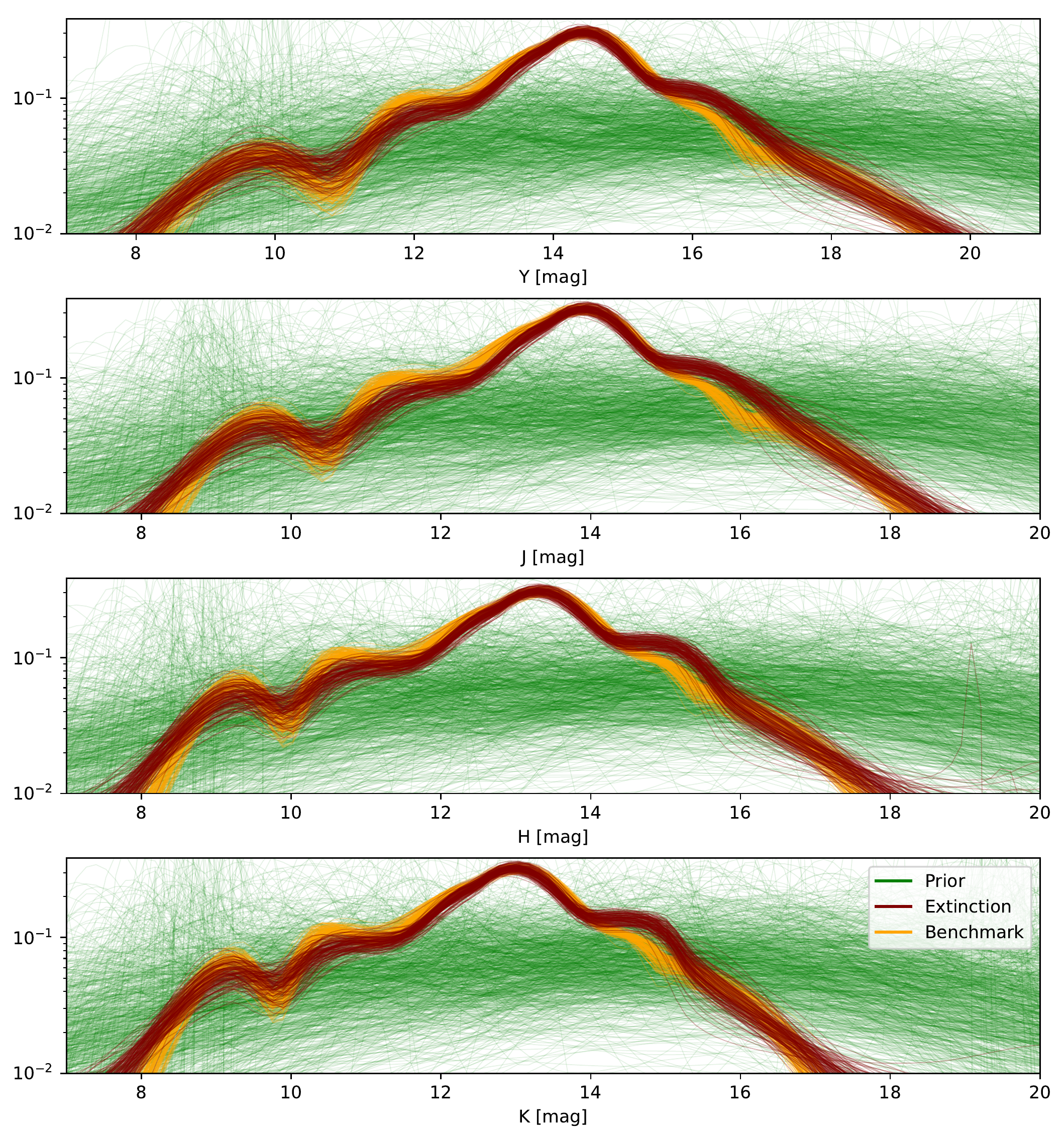}
    }
     \caption{Same as Fig. \ref{fig:comparison_YJHK_2} but using an informative astrometric prior (see text).}
\label{fig:comparison_YJHK_2a}
\end{figure}

In our methodology, the magnitude distributions are obtained by transforming the color index distribution into magnitude distributions by means of the color-index to magnitude relations, which are modeled as spline functions whose coefficients are also inferred from the data (see Paper I for more details). Figures \ref{fig:comparison_YJHK_0} to \ref{fig:comparison_YJHK_2} show samples of the posterior distributions (maroon lines) of the Y, J, H, and K magnitude distributions inferred from the MAP$_{0,1,2}$ data sets. For comparison purposes, the figures also exhibit the magnitude distributions of the benchmark solution (orange lines), as well as the prior (green lines). As can be observed from Fig. \ref{fig:comparison_YJHK_0}, the magnitude distributions inferred from the MAP$_0$ data set are almost identical to those of the benchmark solution, except in the region of K $\sim$ 9 mag where they underestimate the density. Figure \ref{fig:comparison_YJHK_1} shows that the magnitude distributions inferred from the MAP$_1$ data set underestimate the density in the regions of K $\sim$ 9 mag and K $\sim$ 15 mag, and overestimate it at K $\sim$ 10 mag. In addition, the peak of the magnitude distribution of the J band is shifted $\sim$0.5 mag towards the faint end. Finally, magnitude distributions inferred from the MAP$_2$ data set show all the previous effects, together with a loss of precision in the bright end, from 8 mag to 11 mag in the Y band. This latter effect is a direct consequence of the loss of precision in the color index distribution discussed in the previous section. 

This analysis shows that inferred magnitude distributions exhibit artifacts whose magnitude increases with the extinction value of the data set. These artifacts are the reflection of the smoothing of the color index distribution (see Sect. \ref{subsection:parameters}) amplified by the shifts in the coefficients of the spline functions that model the color index to magnitude relations. As discussed in the previous section, the smoothing and shifting of the model parameters have their origin in the contaminants introduced into the model. Although we developed strategies to minimize these contaminants when our methodology is used as a classifier, their influence in the model parameters still impacts the inferred magnitude distributions. Nonetheless, the magnitude distributions inferred from data sets with up to $A_v\sim3$ mag of extinction are recovered with minimal errors. In particular, the magnitude distribution of the K band, the most infrared of our bands, shows no systematic errors.

Because the majority of the contaminants come from sources with missing values (see Sect. \ref{subsection:classifier} and Appendix \ref{appendix:classifier}) the minimization of their impact in the population distributions and in particular in the magnitude distributions requires constraining information able to counterbalance the lack of it resulting from the high percentage of sources
with missing values. The only sources of information in our model are the data set, the extinction map, and the prior distribution. From the statistical modeling point of view, the information content of both the data set and the extinction map is fixed and cannot be improved until the arrival of new and more constraining data.

Luckily, our Bayesian methodology provides a straightforward solution to the low-information problem: the prior distribution. The lack of discriminant information originating from the sources
with missing values can be counterbalanced by the information content of the prior. Thus, taking advantage of the Bayesian formalism, we provide the model with more constraining prior distributions. To avoid biasing the magnitude distributions by the prior itself, we only modify the prior distribution of the astrometric parameters.

Given that the results of the MAP$_2$ data set are the most affected by the lack of constraining information, we reanalyze these data with the following modifications.
We modify the hyper-parameter $\alpha_{sgl}$ of the Dirichlet distribution acting as prior for the proper motion GMM fractions (i.e., weights). Until now, we have used the weakly informative value of $\alpha_{sgl}=[5,4,1,1]$ (see Appendix \ref{subsection:hyper-parameters}); now, we set it to the more constraining value of $\alpha_{sgl}=[50,40,10,1]$. In addition, we modify the hyper-parameters of the Gamma distribution acting as prior of the standard deviations of the GMM. So far, we have used the value of $\beta_{sgl}=[2,2,2,2]\ \ \rm{mas \cdot yr^{-1}}$ (see Appendix \ref{subsection:hyper-parameters}), but now we replace it by $\beta_{sgl}=[4.5, 1.7, 11.9, 52.8] \ \ \rm{mas \cdot yr^{-1}}$, which correspond to the values of the standard deviations found after fitting a GMM to the candidate members reported in Paper I.

The results of our analysis of the MAP$_2$ data set with a more constraining astrometric prior show that the systematic errors in the magnitude distributions are considerably reduced. Figure \ref{fig:comparison_YJHK_2a} shows the magnitude distributions recovered by our methodology using the constraining astrometric priors described above. As shown by this figure, the systematic shift in the J band magnitude distribution is now removed. In addition, the dispersion present at the bright end of the distributions, shown as the dispersed lines between magnitudes 8 and 11 in the K band of Fig. \ref{fig:comparison_YJHK_2}, is now considerably reduced, as shown by the more crowded lines of Fig. \ref{fig:comparison_YJHK_2a} in the same magnitude interval. Finally, although the over-density at K$\sim$15 mag persists, the discrepancy  is nonetheless negligible. On the other hand, the restrictive astrometric prior has negative consequences when our methodology is used as a classifier. Although the contamination rate remains low at 8.7\%, the true positive rate drops by $\sim$20\% to a value of 62.5\% (compare to the value in the last row of Table \ref{table:quality_map_2}).

In this section, we shown that the population distributions inferred from data sets with extinction values up to $A_v\sim3$ mag show only a general broadening resulting from contaminants. Up to 70\% of these contaminants have their origin in the high fraction of missing values of our data set. Moreover, the K band magnitude distribution inferred from these data sets shows no systematic errors. The magnitude distributions inferred from data sets with high extinction ($A_v\sim6$ mag) and a high fraction of sources with  missing values (>96\%) show biases that are minimized thanks to an informative astrometric prior. Nonetheless, this more informative prior results in a $\sim$20\% drop in the true positive rate, which points to the importance of acquiring data sets with fully observed photometric features, particularly in highly extincted regions with $A_v\gtrsim6$ mag.

\section{Conclusion}
\label{section:conclusions}

In this work, we improve the methodology of Paper I by extending its application to embedded stellar clusters and validate it using synthetically extincted data sets of the Pleiades cluster. In the interval of extinctions analyzed here ($A_v\in[0,6]$ mag), our new methodology delivers lists of candidate members with true positive and contamination rates greater than 83\% and lower than 9\%, respectively. This low value of the contamination rate was achieved thanks to a strategy that obtains the probability classification threshold through a detailed analysis of the classifier quality at different extinction bins. The variations at these bins can reach up to 16\% in the contamination rate and down to 70\% in the true positive rate. The application of this strategy to real stellar clusters will require the generation of carefully crafted synthetically extincted data sets that mimic the properties of the real one. 

The inferred magnitude distributions display artifacts whose magnitude increases with the extinction of the data set. For extinction values up to $A_v\sim3$ mag, as those exhibited by the bulk of members in the star-forming regions of Corona Australis, Taurus, and Upper Scorpius (see Fig. \ref{fig:cumulatives}), the artifacts in the K band magnitude distribution are minimal. In data sets with higher extinction values ($A_v \sim 6$ mag), the artifacts can be minimized by increasing the information content of the model, that is, information regarding the data set, the extinction map, and the prior distributions. Given a fixed data set and its extinction map, we demonstrate that the magnitude distributions inferred under a constraining astrometric prior display minimal biases. However, the use of this constraining prior for the inference of population distributions from data sets with a high fraction of sources with missing values  results in a 20\%  drop in the true positive rate of the classifier. Therefore, the decision of whether or not to use  a constraining prior must be motivated by the scientific objective and the information content of the data set.
    
We developed the present methodology to infer the population distributions of embedded young stellar clusters, like those shown in Fig. \ref{fig:cumulatives}. The application of this new methodology to real and extincted stellar clusters will require detailed analyses of the input information provided to the model. The input data set, the extinction map, and the prior distributions must be carefully chosen.

Concerning the data set, we observe that the representation space is crucial. A large number of photometric features implies, in principle, more information to constrain the source extinction. However, the presence of missing values can hamper the performance of our methodology. Therefore, the representation space must maximize the number of photometric bands, in particular the infrared ones, but minimize the fraction of  sources with missing values, particularly those with a missing color index. The latter must, in addition, fulfill the requirements specified in Paper I, in particular regarding the injectivity of the splines, which describe the photometric magnitudes as a function of the color index. 

Concerning the extinction map, we observe that as the extinction value increases, the already scarce information provided by sources with missing values becomes more diluted over the wider limits of the extinction marginalization integral. As explained above, the lack of constraining information introduces contaminants into the model. The extinction map can help to mitigate these contaminants by reducing the limits of the marginalization integral of Eq. \ref{eq:extinction}. For example, a 3D extinction map can be used to replace the uniform extinction prior used here. This new extinction prior can be obtained by marginalizing the 3D extinction map with the aid of the 3D shape of the cluster. However, this approach will require that the  cluster members been found iteratively, the cluster 3D shape be inferred, and that these be used to update the extinction prior.

Concerning the prior distributions, we use weakly informative ones. However, if neither the information content of the data set nor that of the extinction map are enough to mitigate the effects of possible contaminants, then more restricting prior distributions can be used. To minimize possible biases introduced in the magnitude distributions by the prior itself, the constraining information must be included in the astrometric prior. Luckily, the \textit{Gaia} mission is a remarkable source for this constraining information because the cluster members found with only the astrometric features can be used to set stronger astrometric prior distributions.

\begin{acknowledgements}
We thank the anonymous referee for the constructive comments. This research has received funding from the European Research Council (ERC) under the European Union’s Horizon 2020 research and innovation program (grant agreement No 682903, P.I. H. Bouy), and from the French State in the framework of the ”Investments for the future” Program, IdEx Bordeaux, reference ANR-10-IDEX-03-02.
The figures presented here were created using Matplotlib \citep{Hunter:2007}.
\end{acknowledgements}

\bibliographystyle{aa} 
\bibliography{mybiblio}
\begin{appendix}
\section{Methodological and computational improvements}
\label{appendix:improvements}
This Appendix describes the methodological and computational improvements to the original methodology of Paper I. Here, we also provide the details of the hyper-parameters of our model.

\subsection{GPU computation}
Our methodology is computationally expensive. The results presented in Paper I took 30 days to run in a computing server with 80 CPUs at 3.5 GHz each. To improve this, we translated the CPU computation of the likelihood into GPU code using the \textit{numba} compiler \citep{numba}. Using the same data set as in Paper I and a computing machine with eight GPUs Nvidia GForce RTX 2080i, our improved computation takes $\sim$10 hr to run, which represents a speed-up factor of 72.

\subsection{Simplified intrinsic photometric dispersion}
In Paper I (see Sect. 2.1.2), the cluster's intrinsic photometric dispersion was modeled with a multivariate Gaussian distribution with a full covariance matrix. This distribution was evaluated in a color index grid with $n=300$ steps, which resulted in a cluster photometric sequence with a hosepipe shape. The same hosepipe shape can be recovered with a simpler photometric model in which the off-diagonal terms of the covariance matrix are neglected (e.g., in a color--magnitude diagram, the hosepipe shape of the cluster sequence can be obtained by overlapping either ellipses or circles). Here, we reduce the model complexity by neglecting the off-diagonal terms of the covariance matrix. However, to obtain the same value of the original photometric likelihood within a tolerance of $10^{-4}$, we have to increase the number of grid evaluations to $n=500$. Although the evaluation of the photometric likelihood has now 60\% more steps, the number of model parameters is reduced by $D_{ph}\cdot(D_{ph}-1)/2$, where $D_{ph}$ is the photometric dimension of the representation space. This simpler model is sampled more efficiently than the original one (i.e., the one with the full covariance matrix), and thus the total computing time is effectively reduced.

\subsection{New prior families}
\label{subsection:new_prior}
In Paper I, we selected a series of prior families that were the best choice according to the literature. After a critical review of our original choices, we update the following family distributions.

Our choice of the Half-Cauchy family as a prior for variance parameters was inspired by the work of \citet{Gelman2006}. However, it has been shown\footnote{\url{https://mc-stan.org/users/documentation/case-studies/weakly_informative_shapes.html}} that the heavy tails of the Cauchy family can seriously hinder the computational performance of Markov chain Monte Carlo (MCMC) samplers, which increases the computational time required to ensure convergence of the sampler. Therefore, we replace the Half-Cauchy family with the Gamma one. The latter is defined over the standard deviations instead of the variances, and following the recommendations of \citet{Chung2013} we parameterize it as Gamma$(\alpha=2,\beta)$, with $\beta$ a hyper-parameter. 

Similarly, we also replace the \citet{Huang2013} prior of the photometric and astrometric covariance matrices with a Gamma one \citep{Chung2013}. The $\beta$ parameters are also model hyper-parameters.

\subsection{Field model}
\label{subsection:field_model}
In Paper I, the astrometric and photometric field models were assumed to be a Gaussian+Uniform mixture model and Gaussian mixture model (GMM), respectively. The parameters of these models were computed with our Expectation-Maximization routines. Here, we compare the results obtained with two additional literature algorithms (one for the astrometric model and another for the photometric one) over the original field population of Paper I. We still select the best number of components based on the Bayesian information criterion (BIC). 

In the case of the astrometric model, we compare both our original algorithm and the Gaussian Mixture routine from \textit{scikit-learn} \citep{scikit-learn} with 1 to 15 components. The BIC chooses models with 7 components for both algorithms. The difference between the likelihood of the sources resulting from these two best models is significantly reduced when the convergence tolerance is $\leq10^{-6}$. Furthermore, the solutions obtained  after running our complete methodology with these two models are indistinguishable given their uncertainties. Nonetheless, the BIC value of the model inferred by our algorithm is slightly lower than that of the model inferred with \textit{scikit-learn}.  

In the case of the photometric model, due to the presence of missing values, we are only able to test the \textit{ExtremeDeconvolution} algorithm \citep{2011AnApS...5.1657B}. However, the models inferred with this algorithm have components with bizarre behaviors like weights that vanish ($<10^{-10}$) and means that are located at negative magnitudes. Although these effects are strongly reduced by increasing the value of the split-and-merge steps \citep[see Appendix B of][]{2011AnApS...5.1657B}, the computing time increases drastically as well. Our tests with 10 split-and-merge steps put the computing time at $\sim$1-5 hr (in a machine with 48 CPUs at 2.1 GHz) for GMM with 10 to 15 components. If we were to explore the full hierarchy of the split-and-merge tree (with $K(K-1)(K-2)/2$ steps, where K is the number of GMM components), then the computing time of a model with 10 components will be more than a week. Thus, we decided to keep fixed the convergence tolerance at $10^{-6}$ and the split-and-merge steps to 10. Then, the BIC chooses the GMM with 13 components. After running our complete methodology with this field model, we obtain increased contamination in the cluster model; $\sim8$\% more than that obtained when the field model is inferred with our original algorithm. Thus, our original algorithm is, to the best of our knowledge, the most suitable option for this particular problem. Nonetheless, it still suffers from the following problem. Increasing the number of components results in proposed solutions with nonpositive-semi-definite covariance matrices for the components with the lowest weights. Although these proposals are rejected, they still increase the computing time ($\sim$30-40\% per added component for models with more than 15 components). Thus, we only tested models with up to 25 components.

In Paper I, the field likelihood of each source was computed by convolving its uncertainty (see Assumption \ref{assumption:gaussian}) with each Gaussian of the astrometric and photometric field models. Here, we do this as well with further improvement: each Gaussian resulting from the previous convolution is truncated to the data domain, which avoids density leakages (i.e., nonvanishing density in unpopulated regions). 

From the previous comparison, we conclude the following. First, our original algorithms are preferred by the BIC because they were particularly tailored to the characteristics of the original data set from Paper I: proper motions truncated to the interval [-100 mas$\cdot$yr$^{-1}$,100 mas$\cdot$yr$^{-1}$], and a vast fraction of sources
with missing values (>96\%, see Table \ref{table:catalog}). Second, pushing our original convergence tolerance from $10^{-5}$ to $10^{-6}$ we obtain an astrometric GMM with 7 components and a photometric GMM with 19 components. Third, the problem of inferring the parameters of multivariate GMM from data sets with a high percentage  of missing values still presents a computational challenge.

\subsection{Synthetic data}
\label{subsection:synthetic_data}
In Paper I, the probability threshold of classification (i.e., between field and cluster) was found using synthetic data sets generated from the inferred field and cluster models. Here, we do this as well but with the following improvements. First, we generate as many synthetic sources as those present in the real data set. In Paper I, due to computation time reasons, we generated synthetic sources based on a model learned from a sample of only $10^4$ sources. Then, we assumed that the probability threshold found for that model was valid for the model learned in the larger $10^5$ data set. Here, we find that this assumption was not entirely correct. The contamination and true positive rates reported in Paper I were optimistic (more details are given in Appendix \ref{appendix:benchmark_solution}). Second, in Paper I, the uncertainties of the synthetic sources were assigned without consideration about the source origin: cluster or field. As the cluster members tend to have better uncertainties than those of the field population\footnote{The membership probability of a source is the ratio of its cluster likelihood to the total likelihood times the cluster prior probability. Therefore, members tend to have narrow uncertainties that give them large values of the cluster likelihood, which are able to overcome the cluster prior probability, which in our data set is $\sim$ 1\%.}, our original approach resulted in less realistic simulations than those obtained by separating the two cases. Here, the uncertainties and masks of missing values of a synthetic source are assigned as those of a randomly chosen real source in the same magnitude bin. For the latter, we use the most observed photometric band in the real data set. Third, in Paper I, the observed values were assumed to correspond to the synthetic values. Here, the synthetic value and its assigned uncertainty are used as the mean and covariance matrix of a multivariate Gaussian distribution from which the synthetic observed values are drawn. The previous improvements result in more realistic synthetic data sets than those of Paper I.

\subsection{Hyper-parameters}
\label{subsection:hyper-parameters}
The hyper-parameters of our hierarchical model are similar to those defined in Paper I, except for the prior families that were modified as part of our methodological improvements (see Sect. \ref{subsection:new_prior}). The set of model hyper-parameters is shown in Table \ref{table:hyperparameters}. In the latter, the subindices $fc$, $sb$, $sgl$, $clr$, $cfs$, and $ph$, stand for field-cluster, singles-binaries, singles, color, coefficients, and photometry, respectively. $\alpha$ and $\beta$ are the hyper-parameters of the Dirichlet and Gamma distributions, which  are the prior distributions of fractions and standard deviations, respectively. $\mu$ and $\sigma$ are the mean and standard deviation hyper-parameters, respectively, of the univariate and multivariate normal distributions used as prior for the mean proper motions ($\mu_{sgl}$ and $\sigma_{sgl}$), the mean color index ($\mu_{clr}$ and $\sigma_{clr}$), and the mean coefficients of the photometric splines ($\mu_{cfs}$ and $\sigma_{cfs}$). Finally, $rg_{clr}$ and $knots$ indicate the interval of the color index and the knots of the photometric splines, respectively. More details are given in Sect. \ref{section:methodology} and Paper I.

\begin{table}
\caption{Model hyper-parameters.}
\label{table:hyperparameters}
	\begin{center}
	\resizebox{\columnwidth}{!}{
		\begin{tabular}{l c c}
		\toprule
			Name & Units & Value \\
		\midrule
			$\alpha_{fc}$ &  & [980,20] \\
			$\alpha_{sb}$ &  & [8,2] \\
			$\alpha_{sgl}$ &  & [5,4,1,1] \\
			$\beta_{sgl}$ & $\rm{mas\cdot yr^{-1}}$ & [[2,2],[2,2],[2,2],[2,2]] \\
			$\mu_{sgl,0}$ & $\rm{mas\cdot yr^{-1}}$ & [16.13, -39.03] \\
			$\mu_{sgl,1}$ & $\rm{mas\cdot yr^{-1}}$ & [16.13, -39.03] \\
			$\mu_{sgl,2}$ & $\rm{mas\cdot yr^{-1}}$ & [16.13, -39.03] \\
			$\mu_{sgl,3}$ & $\rm{mas\cdot yr^{-1}}$ & [16.13, -39.03] \\
			$\sigma_{sgl,0}$ & $\rm{mas^2\cdot yr^{-2}}$ & [[  17.08, -2.63],[ -2.63,  23.65]] \\
			$\sigma_{sgl,1}$ & $\rm{mas^2\cdot yr^{-2}}$ & [[   2.02, -0.58],[ -0.58,   4.06]] \\
			$\sigma_{sgl,2}$ & $\rm{mas^2\cdot yr^{-2}}$ & [[  143.2, 18.54],[ 18.54, 141.67]] \\
			$\sigma_{sgl,3}$ & $\rm{mas^2\cdot yr^{-2}}$ & [[2940.01,319.19],[319.19,2633.76]] \\
			$\alpha_{clr}$ &  & [1,1,1,1,1] \\
			$\beta_{clr}$ & mag & 1.0 \\
			$\mu_{clr}$ & mag & 4.4 \\
			$\sigma_{clr}$ & mag & 3.6 \\
			$rg_{clr}$ & mag & [0.8, 8.0] \\
			$\beta_{ph}$ & mag & [0.1, 0.01, 0.01, 0.01, 0.01] \\
			$knots$ & mag & [0.8, 3.22, 3.23, 5.18, 8.0] \\
			$\sigma_{cfs}$ & mag & [0.5, 0.5, 0.5, 0.5, 0.5, 0.5, 0.5] \\
			$\mu_{cfs,Y}$ & mag & [ 9.30, 10.81, 12.33, 15.06, 18.04, 21.02, 22.79] \\
			$\mu_{cfs,J}$ & mag & [ 8.75, 10.21, 11.66, 14.29, 17.16, 20.02, 21.71] \\
			$\mu_{cfs,H}$ & mag & [ 8.39,  9.78, 11.18, 13.69, 16.43, 19.17, 20.80] \\
			$\mu_{cfs,K}$ & mag & [ 8.40,  9.68, 10.97, 13.30, 15.83, 18.37, 19.87] \\
		\bottomrule
		\end{tabular}
	}
	\end{center}
\end{table}

\section{The basic and benchmark solutions}
\label{appendix:benchmark_solution}
In this Appendix, we describe the results of the basic and benchmark solutions obtained with the original nonextincted data set (see Sect. \ref{section:data}). The basic solution is obtained with the improved basic module (i.e., the methodological and computational improvements described in Appendix \ref{appendix:improvements}). In contrast, the benchmark solution is obtained with the improved basic module plus the simplified model of EMB (see Sect. \ref{subsection:simplified_emb}). Both basic and benchmark solutions use the same field model, which we obtained after fitting a GMM to the field population (see Appendix \ref{subsection:field_model}).

\subsection{Basic solution}
\label{subsection:basic_solution}
The results of applying the improved basic module to the original data set are the following. The optimum probability threshold (found using the first strategy, that of Paper I) is $p_{t}=0.68$. At this probability threshold, the true positive rate (TPR), contamination rate (CR), and accuracy (ACC) of the classifier are TPR=89.8\%, CR=6.9\%, and ACC=99.6\%. Compared to the solution of Paper I, where $p_t=0.84$, the TPR remains similar, the CR is worst by 3\%, and the ACC is 3\% better. These differences are explained by our improved and more realistic modeling of the synthetic data (see Appendix \ref{subsection:synthetic_data}). The basic solution finds 1990 candidate members, of which 1846 are in common with the results of Paper I. The number of candidate members is compatible, within the Poisson uncertainties, with the 1973 candidate members found in Paper I. The new solution rejects 144 of the original candidate members and finds 122 new ones. The new and rejected candidates are randomly distributed over the proper motions and photometric sequence of the cluster and show no evidence of systematic shifts. Moreover, their numbers are compatible with the original ones given the measured value of CR (6.9\%). Therefore, we conclude that this new solution is statistically indistinguishable from that of Paper I.

\subsection{Benchmark solution}
Here, we show the details of the benchmark solution obtained with the basic+EMB modules over the original data set. First, we analyze the quality of the classifier in terms of the TPR and CR, and then we describe the inferred posterior distributions of the proper motions, the color index, and Y, J, H, and K bands.
\subsubsection{Quality of the classifier}
\label{subsubsection:benchmark_quality}
\begin{figure}
    \centering
    \includegraphics[width=\columnwidth,page=1]{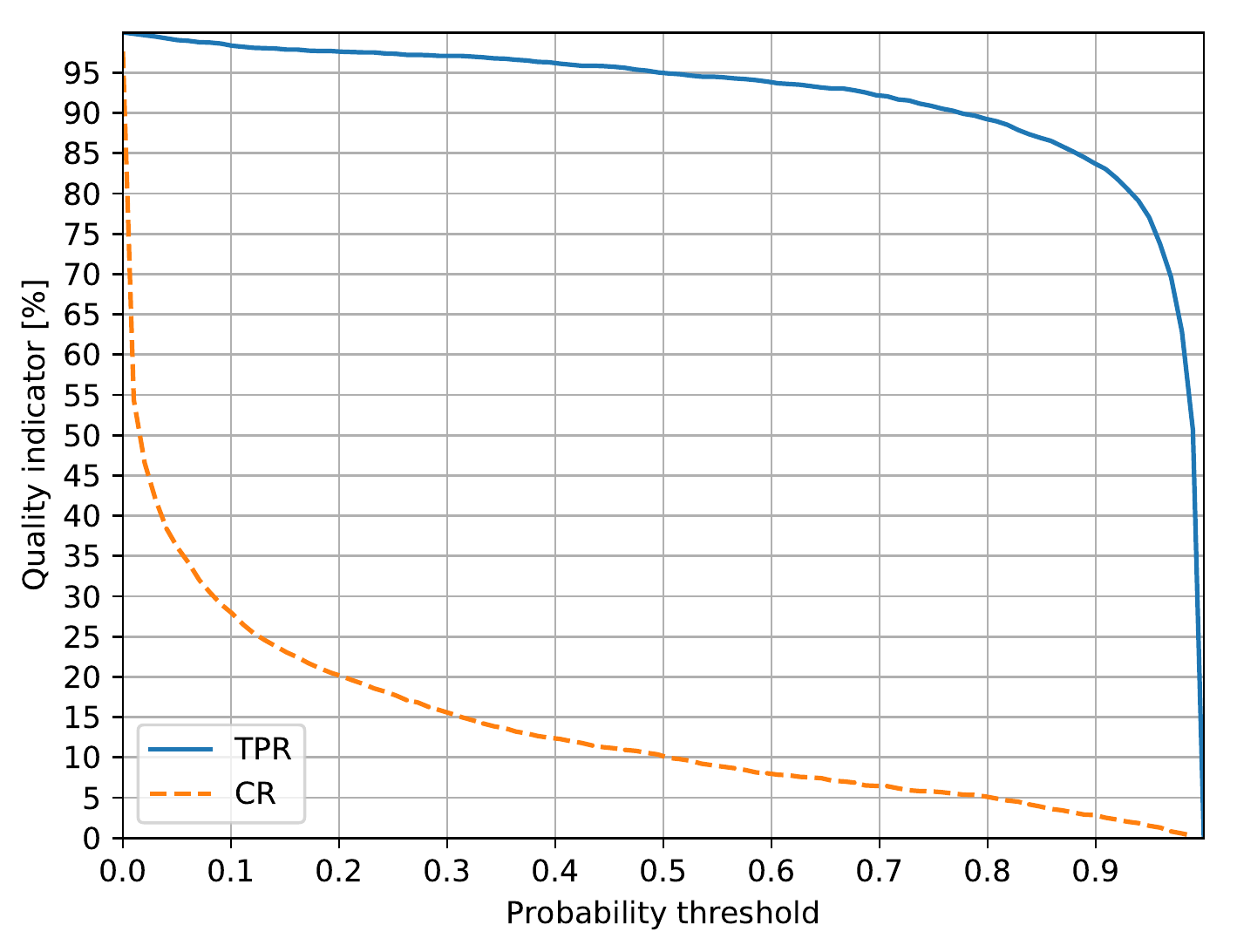}
     \caption{TPR and CR of the benchmark solution as functions of the probability threshold.}
\label{fig:benchmark_quality}
\end{figure}

The benchmark solution finds 1943 candidate members with a membership probability larger than the optimum probability threshold of $p_t=0.68$, hereafter we refer to these as the high membership probability sample (HMPS). Compared to the 1990 candidate members found by the basic solution (see Sect. \ref{subsection:basic_solution}), there are 1903 in common, 87 rejected, and 40 new ones. The quality indicators of the classifier are shown in Fig. \ref{fig:benchmark_quality}, in particular, at the optimum probability threshold, the TPR is 92.5\%, the CR is 6.5\%, and the ACC is 99.6\%.

Although the benchmark solution found 47 less candidate members  than the basic solution, both numbers are compatible within the Poisson uncertainty. Moreover, under the assumption that the members of the basic solution are the true ones, the 1903 common members represent 95.6\% of the 1990 candidate members, well above the 92.5\% of the TPR reported by the benchmark solution. The previous values indicate that our quality measurements are consistent between solutions and that the benchmark solution is similar to the basic one. However, contrary to what is observed when comparing the basic solution to that of Paper I, in this case, the rejected sources are not randomly distributed.

The 87 rejected candidate members are located at a mean distance of 15.1 $\rm{mas\cdot yr^{-1}}$ from the cluster proper motion center, whereas the common candidate members have a mean distance of 5.2 $\rm{mas\cdot yr^{-1}}$, showing that the rejected sources are mainly on the outskirts of the proper motions distributions. Furthermore, 72 out of the 87 rejected sources are farther away than the 5.2 $\rm{mas\cdot yr^{-1}}$ of typical distance in the common candidate members. The rejected sources were considered members in the basic solution due to the more extended wings of the EMB proper motions model, but once this model is removed (see Sect. \ref{subsection:simplified_emb}), the membership probabilities of the rejected sources fall below the optimum probability threshold and are therefore no longer considered as candidate members. Although 30\% of the rejected candidate members from the basic solution are EMB, the total fraction of EMB reported by both solutions is $\sim$14\%. Therefore, we conclude that although our EMB module results in a simplified model of the EMB with respect to that of Paper I, it still recovers the same fraction of EMB that the basic solution.

\subsubsection{Population distributions}
\label{subsubsection:benchmark_distributions}

\begin{figure}[!ht]
    \centering
    \includegraphics[width=\columnwidth,page=1]{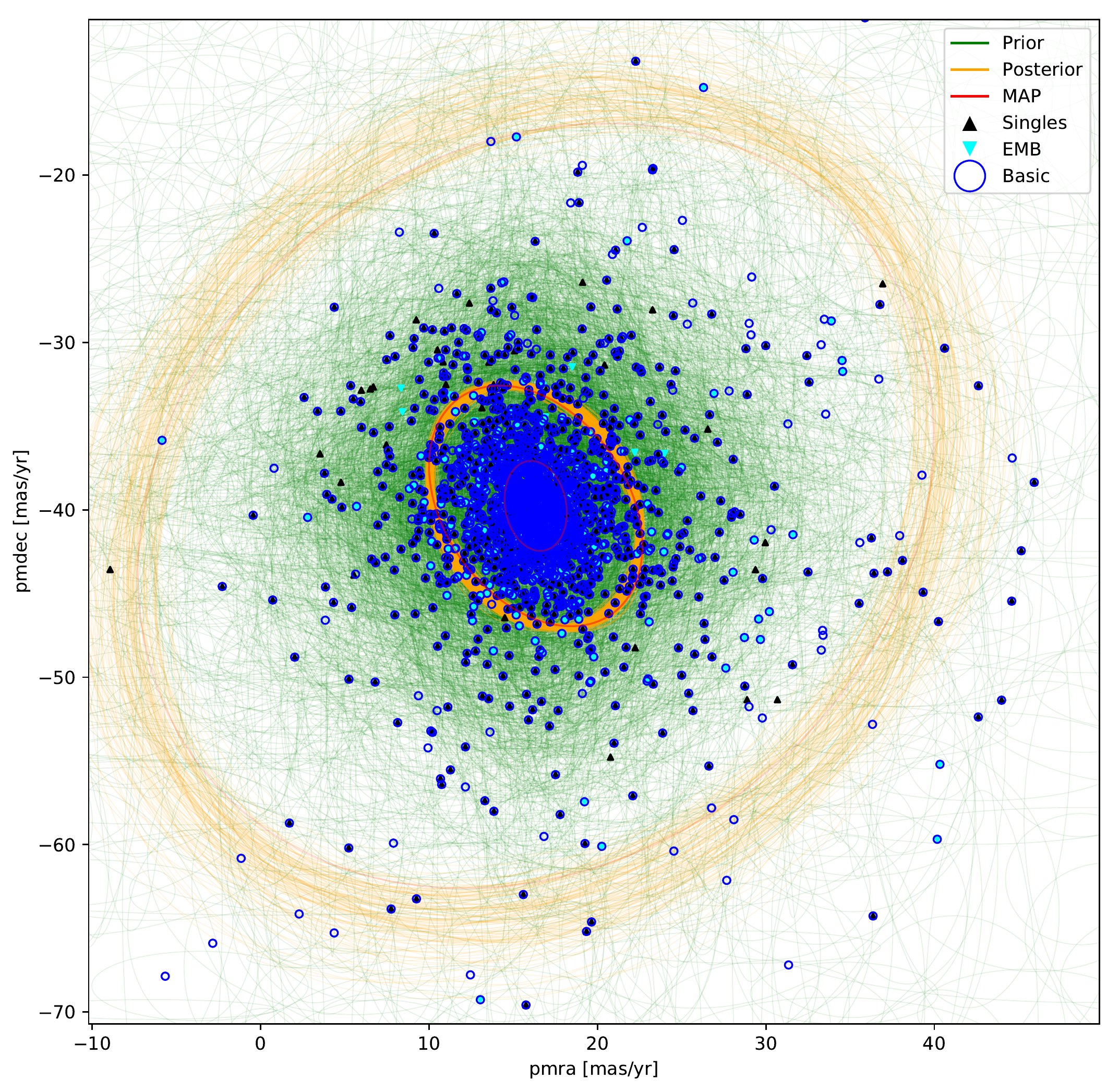}
     \caption{Proper motions diagram showing the single and EMB candidate members found by the benchmark solution, together with samples from the prior (green lines) and posterior (orange lines) distributions of the astrometric GMM parameters, and the maximum-a-posteriori solution (red line). For comparison, the candidate members found with the basic solution are shown with blue circles.}
\label{fig:benchmark_VPD}
\end{figure}

\begin{figure}
    \centering
    \includegraphics[width=\columnwidth,page=2]{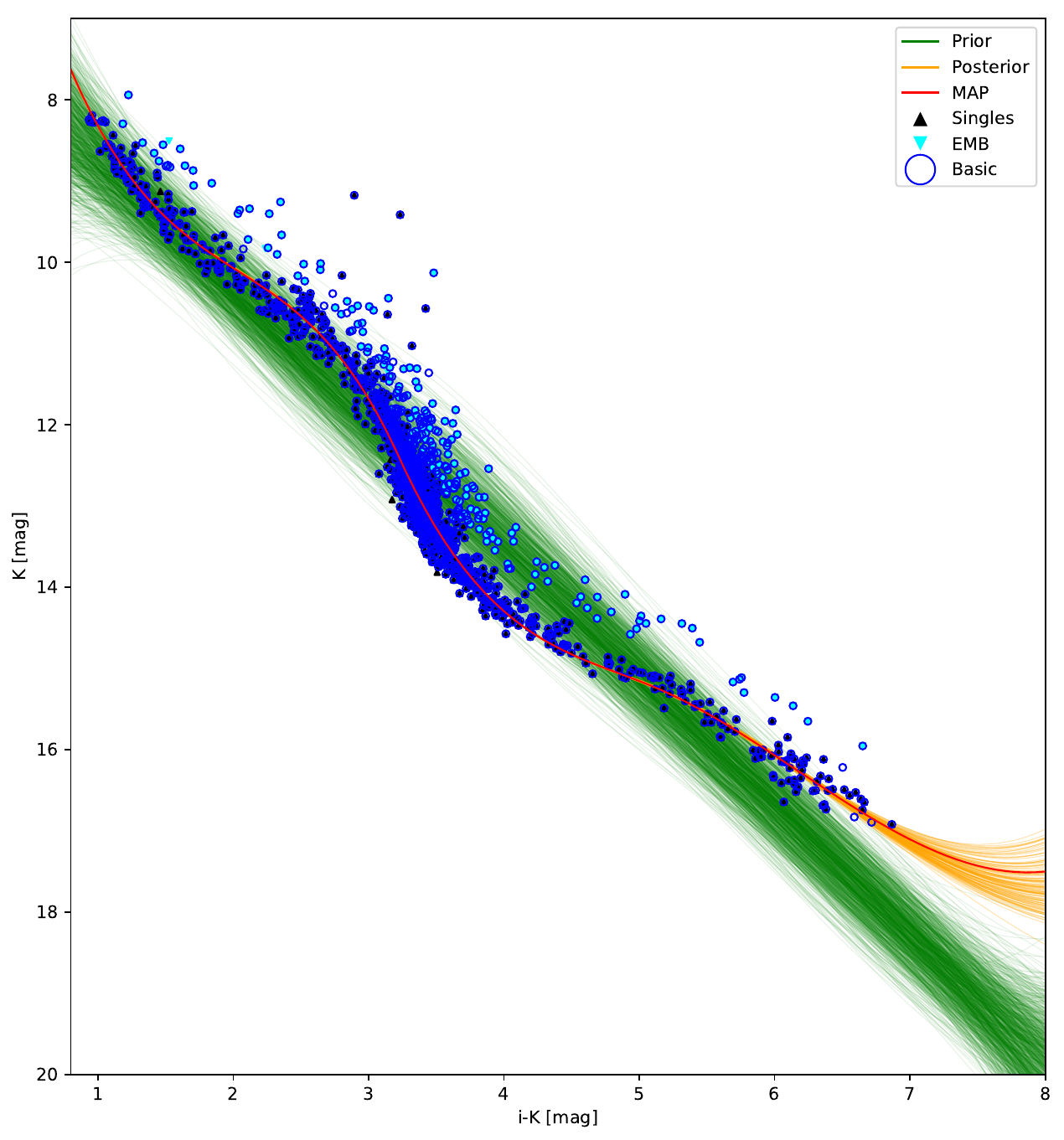}
     \caption{Color--magnitude diagram showing the single and EMB candidate members found by the benchmark solution. Captions as in Fig. \ref{fig:benchmark_VPD}.}
\label{fig:benchmark_CMD}
\end{figure}

\begin{figure}
    \centering
    \includegraphics[width=\columnwidth,page=3]{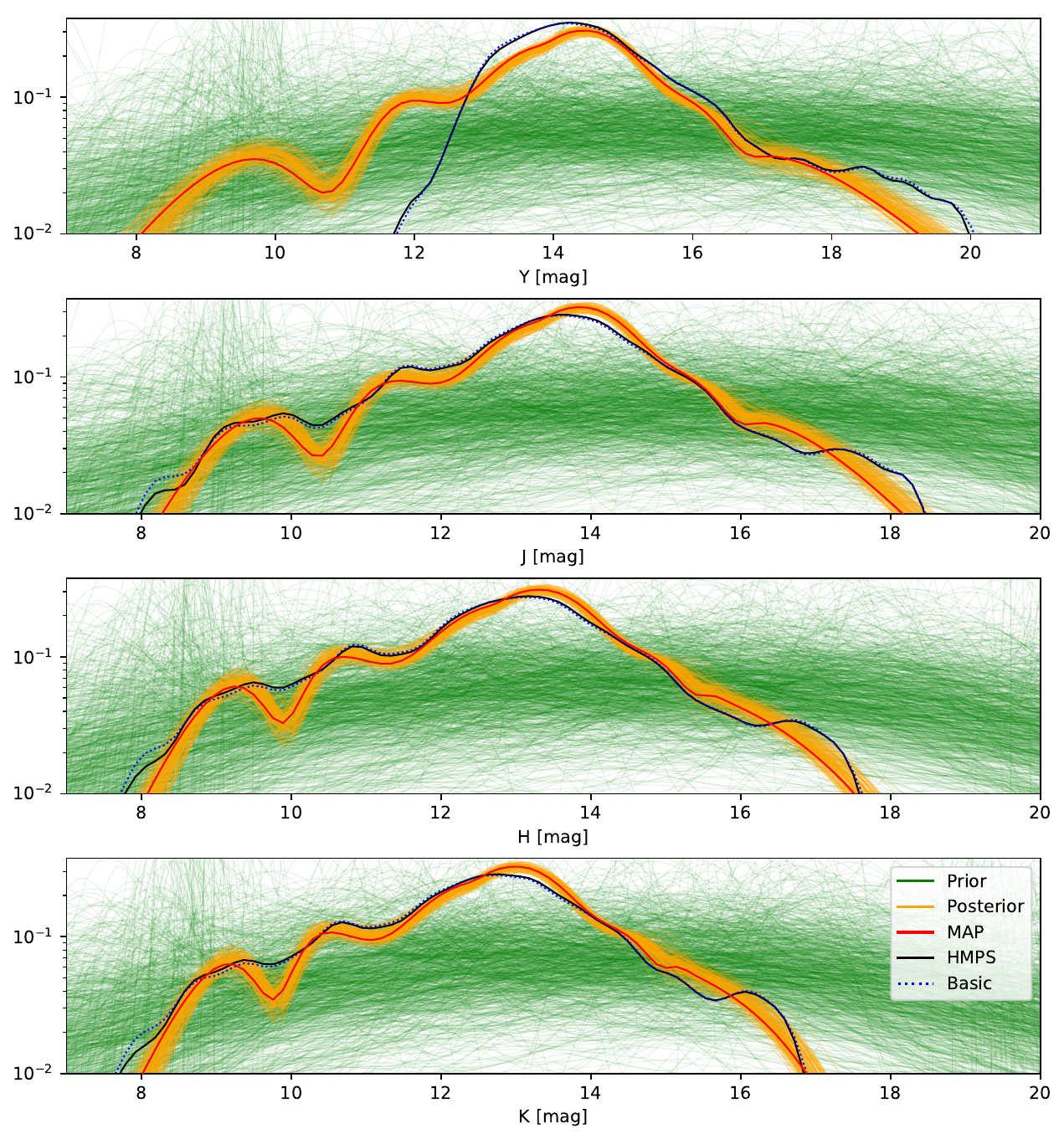}
     \caption{Magnitude distributions in the Y, J, H, and K bands found by the benchmark solution. Captions as in Fig. \ref{fig:benchmark_VPD}.}
\label{fig:benchmark_magnitude}
\end{figure}

We now make a summary of the population distributions found with the benchmark solution. Figures \ref{fig:benchmark_VPD} to \ref{fig:benchmark_magnitude} show the proper motions, color--magnitude, and magnitude diagrams of the HMPS candidate members, together with samples from the prior (green lines) and posterior (orange lines) distributions of the benchmark solution. For comparison, the figures also show the distributions resulting from the candidate members of the basic solution (see Sect. \ref{subsection:basic_solution}). Figure \ref{fig:benchmark_VPD} shows that the majority of the rejected candidate members from the basic solution are located at the outskirts of the proper motion distribution, as already discussed in the previous section. Despite the previous effect, neither the color--magnitude diagram nor the resulting magnitude distributions exhibit any potential bias (see Figs. \ref{fig:benchmark_CMD} and \ref{fig:benchmark_magnitude}). Furthermore, the magnitude distributions resulting from the HMPS are almost indistinguishable from those obtained with the candidate members of the basic solution. The differences between the magnitude distributions obtained with the HMPS and with the posterior distributions of the model parameters are a direct consequence of the fact that the former is not a random sample of the latter, but it only contains the high membership probability sources. In addition, these latter sources have missing values, which prevent us from drawing them, as clearly shown in the bright side of the Y band magnitude distribution.

In this Appendix, we show that the basic and benchmark solutions are almost identical with no evidence of biases. Furthermore, our simplified EMB model recovers the same fraction of EMBs as the basic solution.

\section{Details of the classifier quality}
\label{appendix:classifier}

In this Appendix, we analyze the quality of the classifier obtained using the extinction+EMB modules (see Sect. \ref{subsection:extinction} and \ref{subsection:simplified_emb}) when applied over the synthetically extincted MAP$_{0,1,2}$ data sets. In particular, we measure the accuracy, true positive rate (TPR), and contamination rate (CR) at the optima probability thresholds obtained with the following three strategies.

In the first strategy, we learn the cluster and field models from the synthetically extincted data set using the extinction+EMB module. Then, with these learned models, we generate a new synthetic data set in which the true class labels are known (i.e., those of the field and the cluster). We notice that thanks to our extinction module, the learned cluster model is free of extinction, and therefore the synthetic data generated from it are also extinction-free. Afterward, we infer the cluster model from the previous data using our basic+EMB module. Finally, we compute the optimum probability threshold using the true classes of the synthetic data and the inferred membership probabilities. This strategy follows the approach of Paper I in the sense that it assumes that the learned model is the true one. In the absence of the true labels of the cluster and field populations, this strategy is, to the best of our knowledge, the only available one to obtain the optimum probability threshold. The caveat of it is that the learned cluster model is free of extinction, and therefore the resulting classification threshold is independent of the extinction.

In the second and third strategies, the optimum probability threshold is computed assuming that the true class labels correspond to those obtained by the benchmark solution (i.e., that found with our basic+EMB module on the original nonextincted data set; see Appendix \ref{appendix:benchmark_solution}). While the second strategy computes the optimum probability threshold using all sources independently of their extinction value, the third strategy splits the data into bins of one magnitude of extinction and obtains optimal classification thresholds for each bin. This latter strategy gives a more detailed analysis of the performance of our extinction methodology under different degrees of extinction. The caveat of these strategies is that they assume that the data set contains the true class labels.

Tables \ref{table:quality_map_0} to \ref{table:quality_map_2} show the true positive rate (TPR) and contamination rate (CR) measured with our three strategies on the synthetic data sets MAP$_{0,1,2}$. The rows of these tables show the results of the strategies and are marked as follows. The results of the first and second strategies are shown in the first two rows, while the rest of the rows show those of the third strategy. This latter is shown at each extinction bin and for the concatenation of the labels in all bins (last row). The columns show the average extinction value of each bin, the optimum probability threshold $p_t$, and the quality indicators of the classifier. These latter are shown for the entire data set (labeled All) and for subsets containing only the sources with observed and missing color index (labeled CI$_{obs}$ and CI$_{na}$, respectively). As can be observed, our strategies report high and similar values of the true positive rate. On the contrary, our second and third strategies report contamination rates that are lower than that obtained with the first strategy. The differences between the contamination and true positive rates of the first and second strategies result from their different probability thresholds. As the latter decreases, more members are recovered, but also more contaminants are included. For this reason, the first strategy reports larger true positive and contamination rates than those of the second strategy. As mentioned above, the first two strategies obtain the optimum probability threshold using all sources from the data set. However, thanks to the partitioning of the data into bins of extinction, the third strategy obtains probability thresholds that are optimal for each extinction bin. These thresholds monotonically decrease with increasing extinction as a consequence of both the optimization of the recovered members and the dilution of information across the increasingly wider limits of the marginalization integral of Eq. \ref{eq:extinction}. The previous analysis shows that while our first strategy is similarly good at recovering the cluster members as the other two, the detailed analysis in bins of extinction performed by the third one reduces the contamination rate to values $\lesssim$10\%.

\begin{table}[ht!]
\caption{Quality indicators found in the MAP$_0$ data set.}
\label{table:quality_map_0}
\centering
\resizebox{\columnwidth}{!}{
\begin{tabular}{c|c|c|ccc|ccc|}
\toprule
{} & $A_v$ & $p_t$ & \multicolumn{3}{|c|}{TPR [\%]} & \multicolumn{3}{|c|}{CR [\%]} \\
{} &       &       &     All & CI$_{obs}$ & CI$_{na}$ &    All & CI$_{obs}$ & CI$_{na}$ \\
Strategy &       &       &         &            &           &        &            &           \\
\midrule
First    &     - &  0.70 &   96.30 &      96.89 &     95.01 &  10.39 &       3.79 &     22.14 \\
Second   &     - &  0.79 &   93.90 &      95.94 &     89.50 &   5.89 &       2.98 &     12.01 \\
Bin 1    &  0.10 &  0.79 &   96.32 &      98.03 &     92.66 &   5.39 &       2.88 &     10.62 \\
Bin 2    &  1.71 &  0.67 &   67.48 &      71.26 &     58.33 &  16.16 &       4.62 &     38.24 \\
All bins &     - &     - &   94.39 &      96.18 &     90.53 &   5.97 &       2.97 &     12.19 \\
\bottomrule
\end{tabular}

}
\end{table}

\begin{table}[ht!]
\caption{Quality indicators found in the MAP$_1$ data set.}
\label{table:quality_map_1}
\centering
\resizebox{\columnwidth}{!}{
\begin{tabular}{c|c|c|ccc|ccc|}
\toprule
{} & $A_v$ & $p_t$ & \multicolumn{3}{|c|}{TPR [\%]} & \multicolumn{3}{|c|}{CR [\%]} \\
{} &       &       &      All & CI$_{obs}$ & CI$_{na}$ &     All & CI$_{obs}$ & CI$_{na}$ \\
Strategy &       &       &          &            &           &         &            &           \\
\midrule
First    &     - &  0.75 &    89.55 &      92.04 &     84.17 &   18.00 &       6.32 &     36.66 \\
Second   &     - &  0.87 &    83.45 &      87.74 &     74.18 &    8.91 &       3.50 &     20.33 \\
Bin 1    &  0.61 &  0.95 &    90.06 &      94.34 &     82.50 &    4.78 &       1.96 &     10.00 \\
Bin 2    &  1.52 &  0.90 &    87.81 &      92.33 &     78.72 &    7.28 &       2.24 &     17.32 \\
Bin 3    &  2.66 &  0.64 &    86.47 &      89.79 &     78.39 &   10.77 &       3.70 &     25.95 \\
All bins &     - &     - &    87.53 &      91.32 &     79.35 &    8.64 &       2.96 &     20.24 \\
\bottomrule
\end{tabular}

}
\end{table}

\begin{table}[ht!]
\caption{Quality indicators found in the MAP$_2$ data set.}
\label{table:quality_map_2}
\centering
\resizebox{\columnwidth}{!}{
\begin{tabular}{c|c|c|ccc|ccc|}
\toprule
{} & $A_v$ & $p_t$ & \multicolumn{3}{|c|}{TPR [\%]} & \multicolumn{3}{|c|}{CR [\%]} \\
{} &       &       & All & CI$_{obs}$ & CI$_{na}$ &    All & CI$_{obs}$ & CI$_{na}$ \\
Strategy &       &       &         &            &           &        &            &           \\
\midrule
First    &     - &  0.72 &   84.76 &      87.10 &     79.69 &  22.27 &       7.60 &     43.47 \\
Second   &     - &  0.87 &   77.41 &      82.96 &     65.40 &   9.48 &       3.70 &     22.29 \\
Bin 1    &  0.31 &  0.92 &   86.50 &      92.58 &     73.61 &   5.66 &       2.30 &     13.59 \\
Bin 2    &  1.48 &  0.90 &   85.71 &      93.24 &     71.30 &   9.70 &       4.46 &     20.62 \\
Bin 3    &  2.48 &  0.74 &   85.71 &      91.21 &     74.73 &  13.65 &       5.14 &     29.17 \\
Bin 4    &  3.47 &  0.33 &   84.25 &      86.34 &     78.87 &   9.32 &       3.07 &     23.29 \\
Bin 5    &  4.48 &  0.08 &   77.78 &      78.05 &     77.19 &   7.89 &       2.04 &     18.52 \\
Bin 6    &  5.50 &  0.03 &   70.21 &      70.87 &     68.42 &   9.17 &       8.75 &     10.34 \\
All bins &     - &     - &   83.83 &      88.38 &     74.01 &   8.61 &       3.65 &     19.32 \\
\bottomrule
\end{tabular}

}
\end{table}

\begin{figure}[ht!]
    \centering
    \includegraphics[width=\columnwidth,page=3]{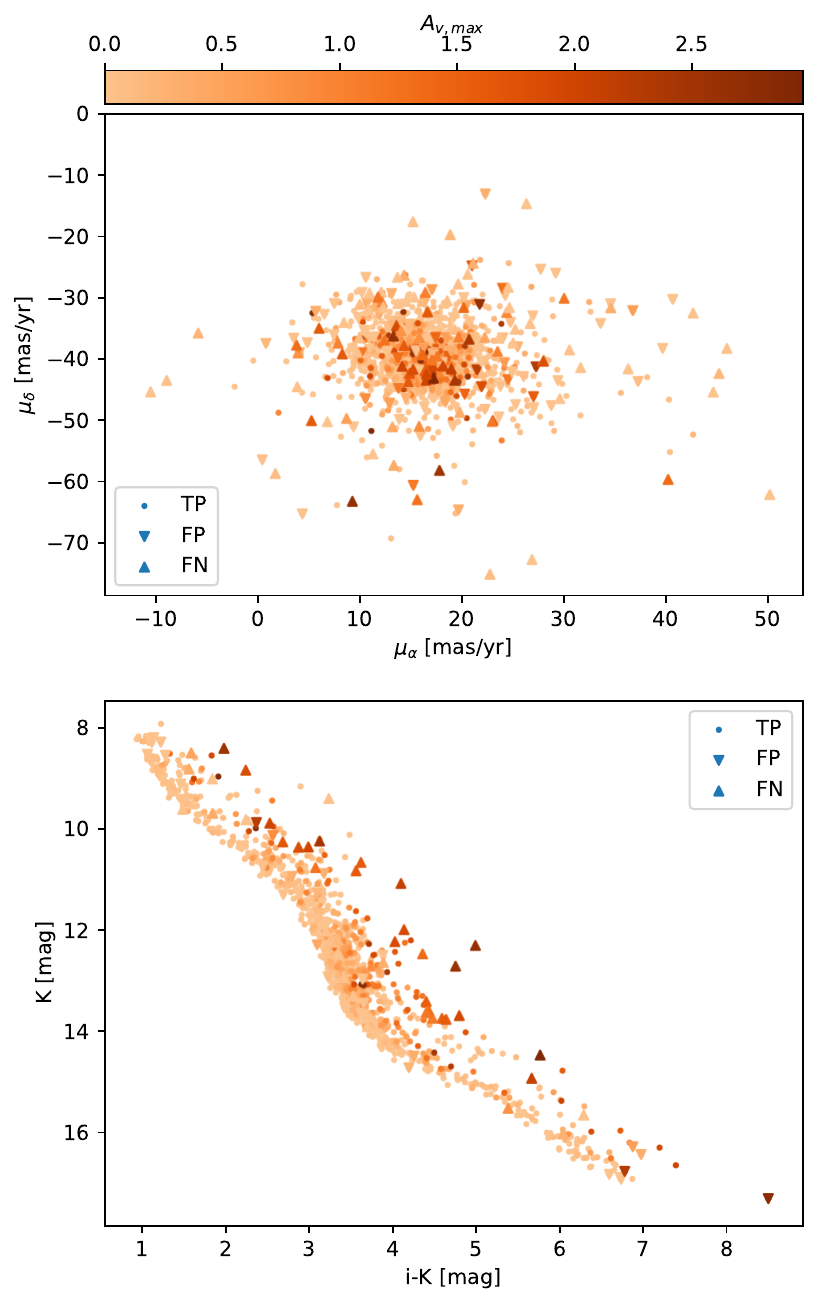}
     \caption{Proper motions and color--magnitude diagrams showing the true positives, false positives, and false negatives, from the MAP$_0$ data set.}
\label{fig:members_map_0}
\end{figure}

\begin{figure}[ht!]
    \centering
    \includegraphics[width=\columnwidth,page=3]{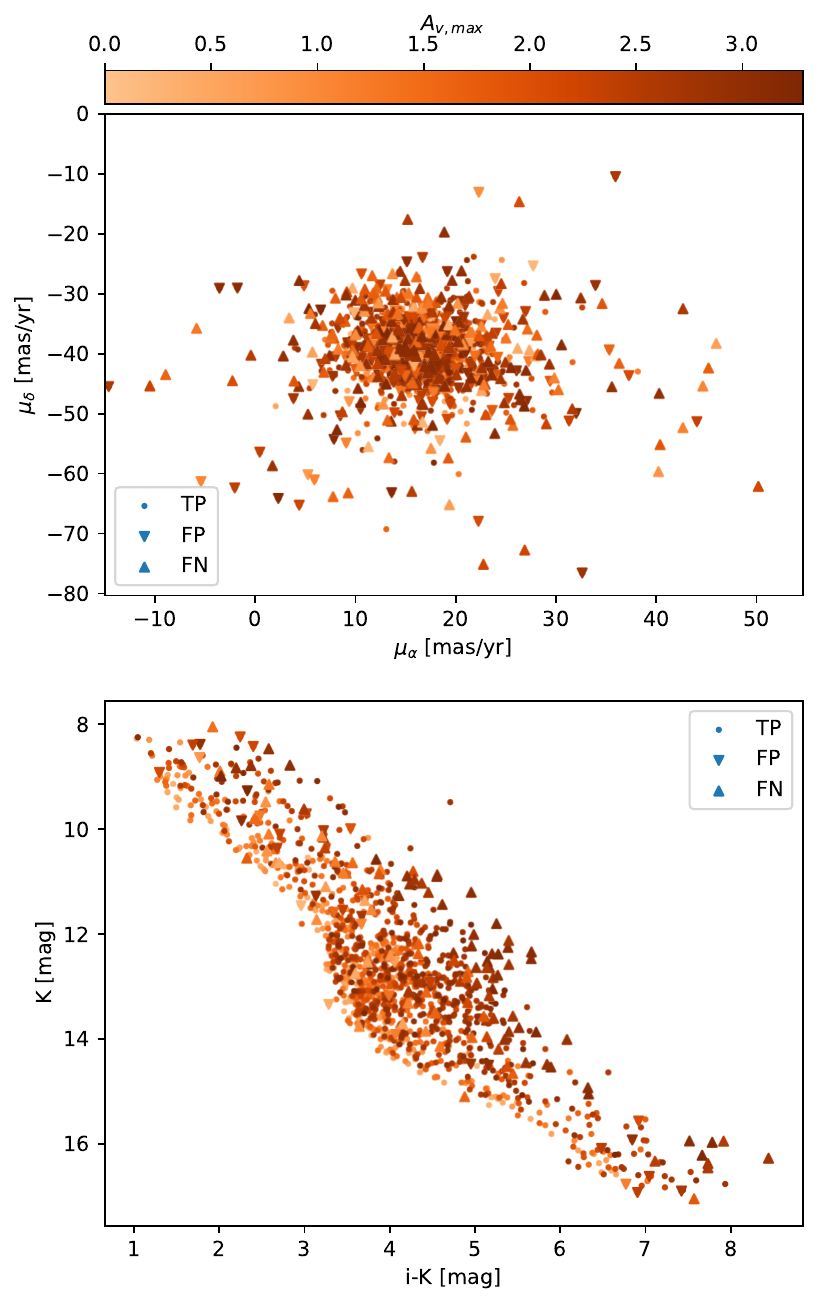}
     \caption{Same as Fig. \ref{fig:members_map_0} but for the MAP$_1$ data set.}
\label{fig:members_map_1}
\end{figure}

\begin{figure}[ht!]
    \centering
    \includegraphics[width=\columnwidth,page=3]{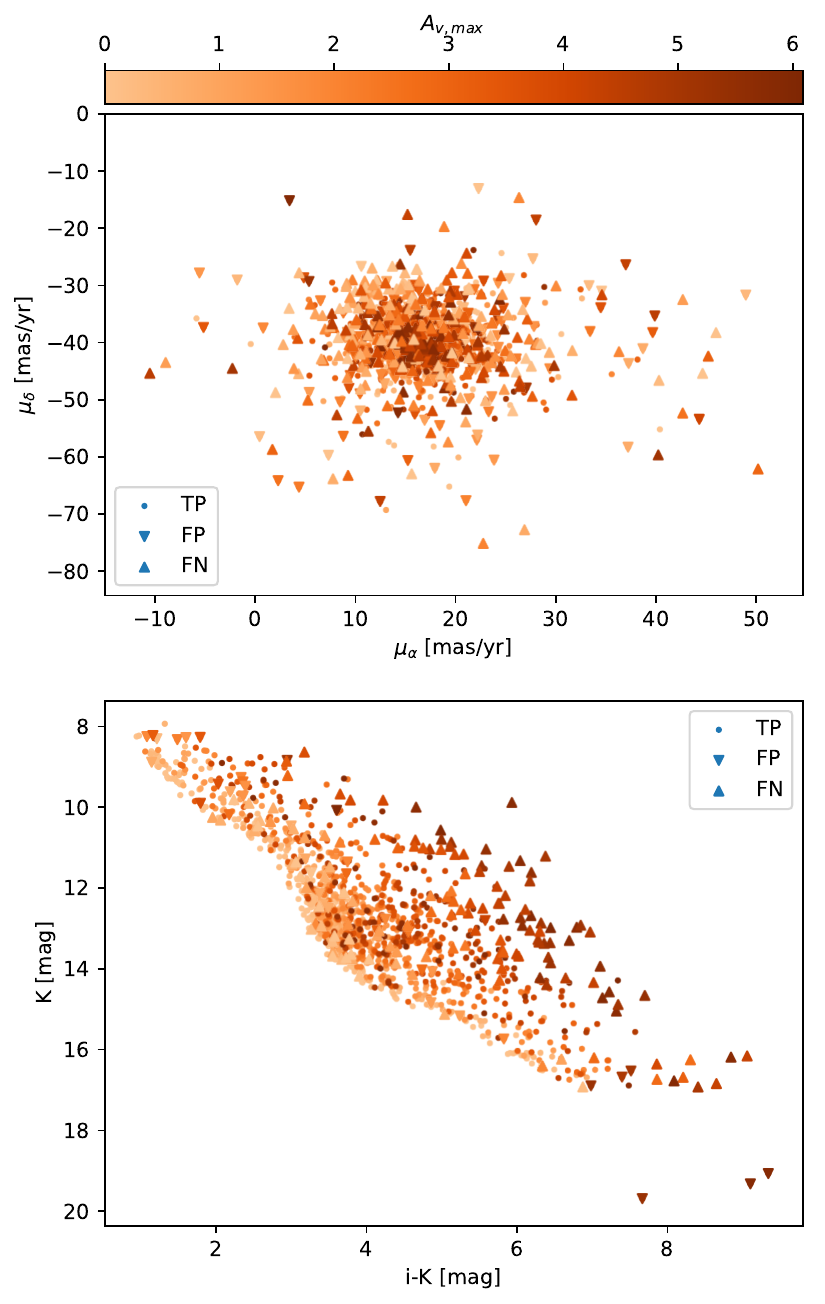}
     \caption{Same as Fig. \ref{fig:members_map_0} but for the MAP$_2$ data set.}
\label{fig:members_map_2}
\end{figure}

Figures \ref{fig:members_map_0} to \ref{fig:members_map_2} show the proper motions and color--magnitude diagrams of sources classified as true positives, false positives, and false negatives in the MAP$_{0,1,2}$ data sets as recovered with the optimum probability thresholds of the third strategy. As can be observed from the proper motions diagram, the false positives and false negatives appear to be more frequent in the outskirts of the distribution than the true positives. On the other hand, the color--magnitude diagrams show that the false negatives tend to have large extinction values, while the false positives are evenly spread in extinction and follow the color--magnitude distribution of the true positives. Therefore, we conclude that our extinction methodology has difficulties in recovering the most extincted members (i.e., the false negatives with $A_v\sim6$ mag), particularly those at the outskirts of the distribution of proper motions, where the confusion with the field population increases. Nonetheless, even at the high extinction value of $A_v\sim5$ mag the true positive rate and contamination rates are better than 70\% and 10\%, respectively.

As discussed in Paper I and confirmed by the values shown in Tables \ref{table:quality_map_0} to \ref{table:quality_map_2}, the sources with missing values, particularly those with a missing color index, are the main culprit of the contamination reported by both our original and extinction methodologies. In Paper I, we observed that sources with a missing color index increased the contamination rate by a negligible $\sim2$\%. Here, we observe that $\sim$70\% of the false positives (i.e., contaminants) come from sources with a missing color index, which unfortunately is the majority of the sources in our original data set (see Table \ref{table:catalog}). Although the contamination rate introduced by these sources in the high extinction regions ($A_v > 5$ mag) of the cluster is $\sim$10\%, the majority of the contaminants come from field sources with missing values that were originally at the blue side of the photometric sequence, but that after being reddened are now compatible with the photometric model thanks to extinction values within the limits of the marginalization integral (see Eq. \ref{eq:extinction}). Thus, the rise in the number of contaminants is a direct consequence of our new and more complex extinction methodology in combination with the lack of discriminant information: the color index.

\subsection{Quality as a function of extinction and magnitude}
In this section, we provide further details of the quality of the  classifier as a function of extinction and magnitude. As the first strategy obtains a cluster model free of extinction, we only analyze its results as functions of magnitude.

\begin{figure}
    \centering
    \includegraphics[width=\columnwidth,page=1]{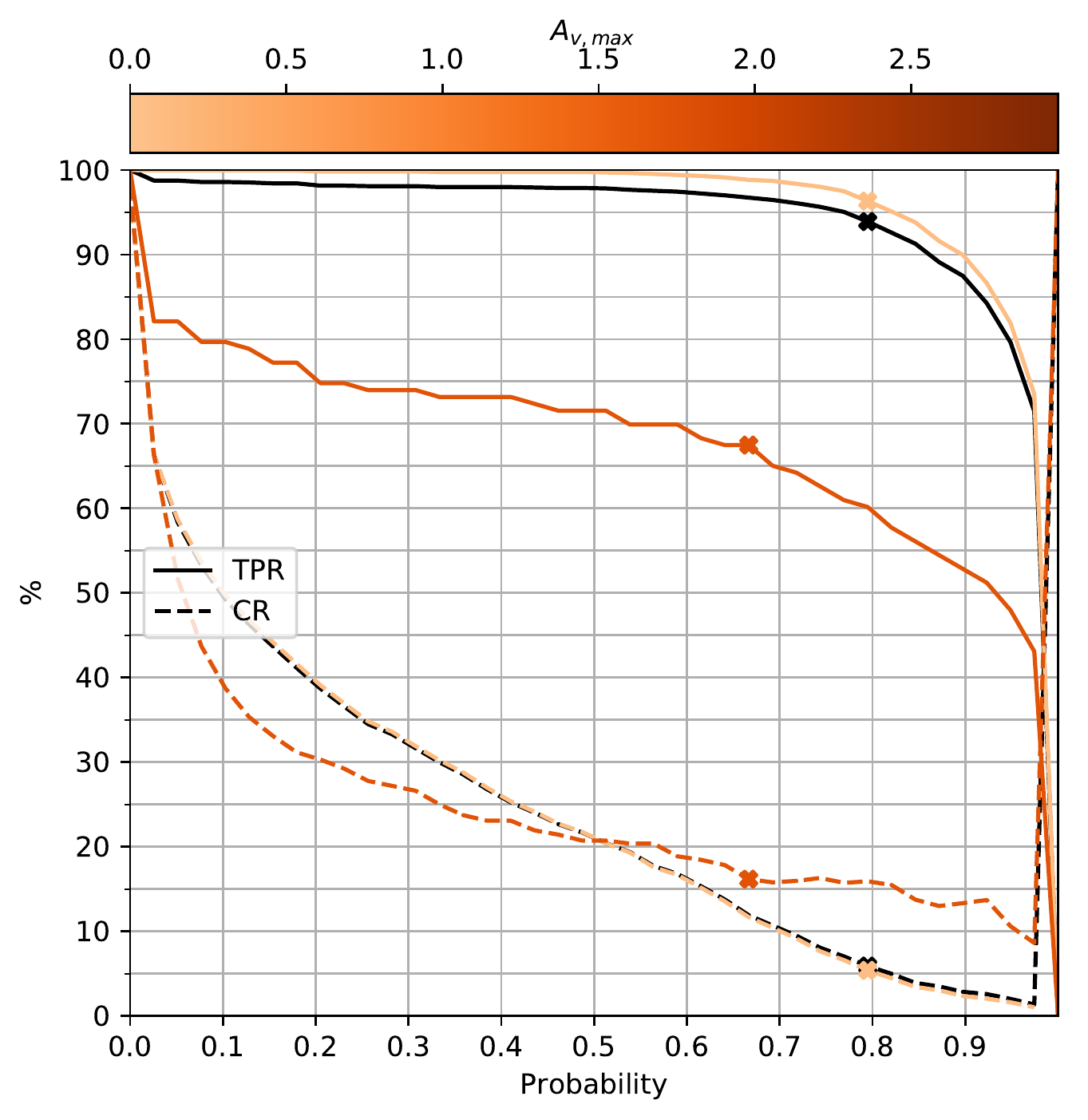}
     \caption{Quality indicators of the classifier obtained from the MAP$_0$ data set. The TPR and CR of the classifier are shown 
     as functions of the probability threshold (black line) and for bins of one magnitude of extinction (color lines).}
\label{fig:quality_map_0}
\end{figure}

\begin{figure}
    \centering
    \includegraphics[width=\columnwidth,page=1]{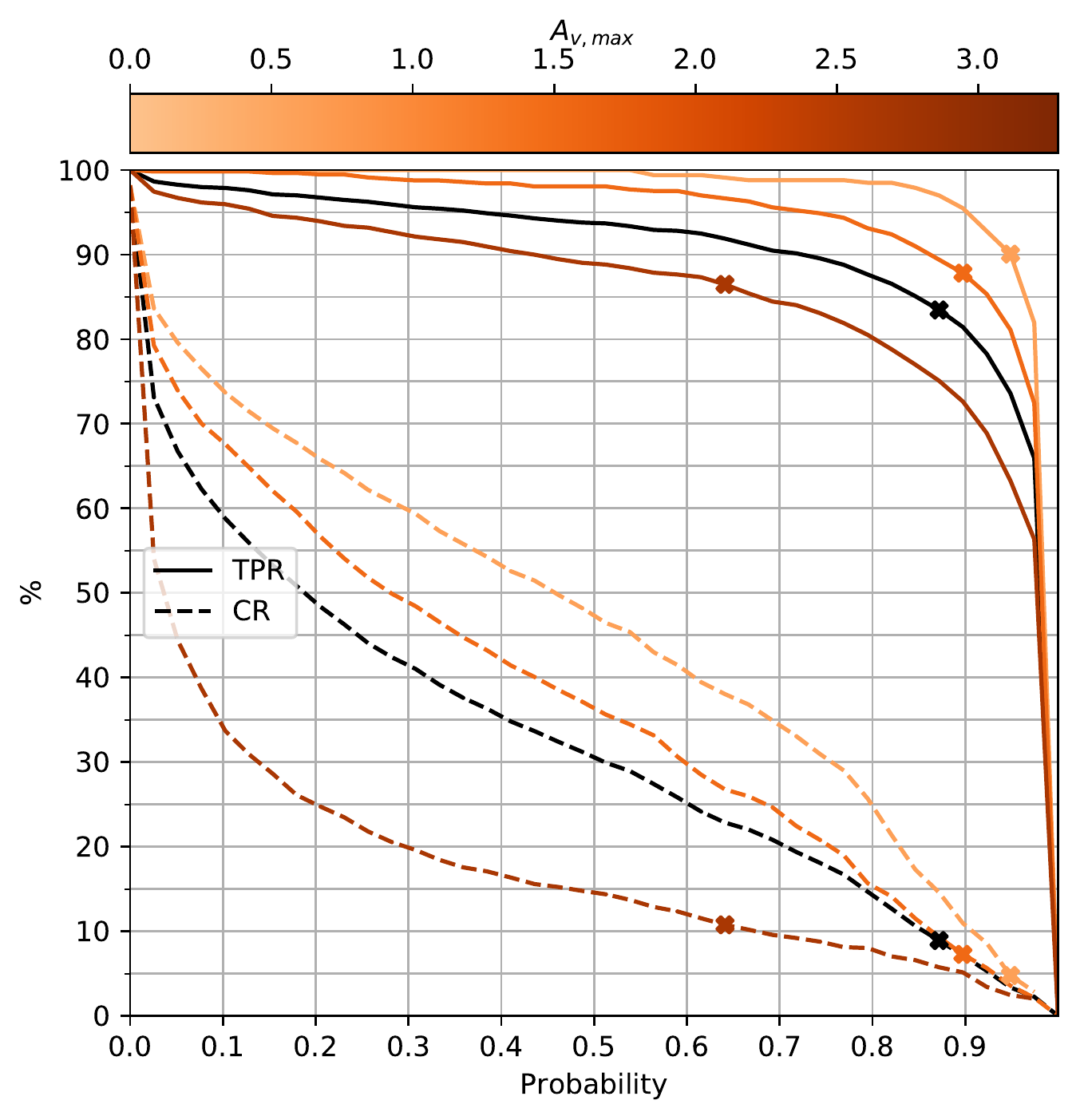}
     \caption{Quality indicators of the classifier obtained the MAP$_1$ data set. Captions as in Fig. \ref{fig:quality_map_0}.}
\label{fig:quality_map_1}
\end{figure}

\begin{figure}
    \centering
    \includegraphics[width=\columnwidth,page=1]{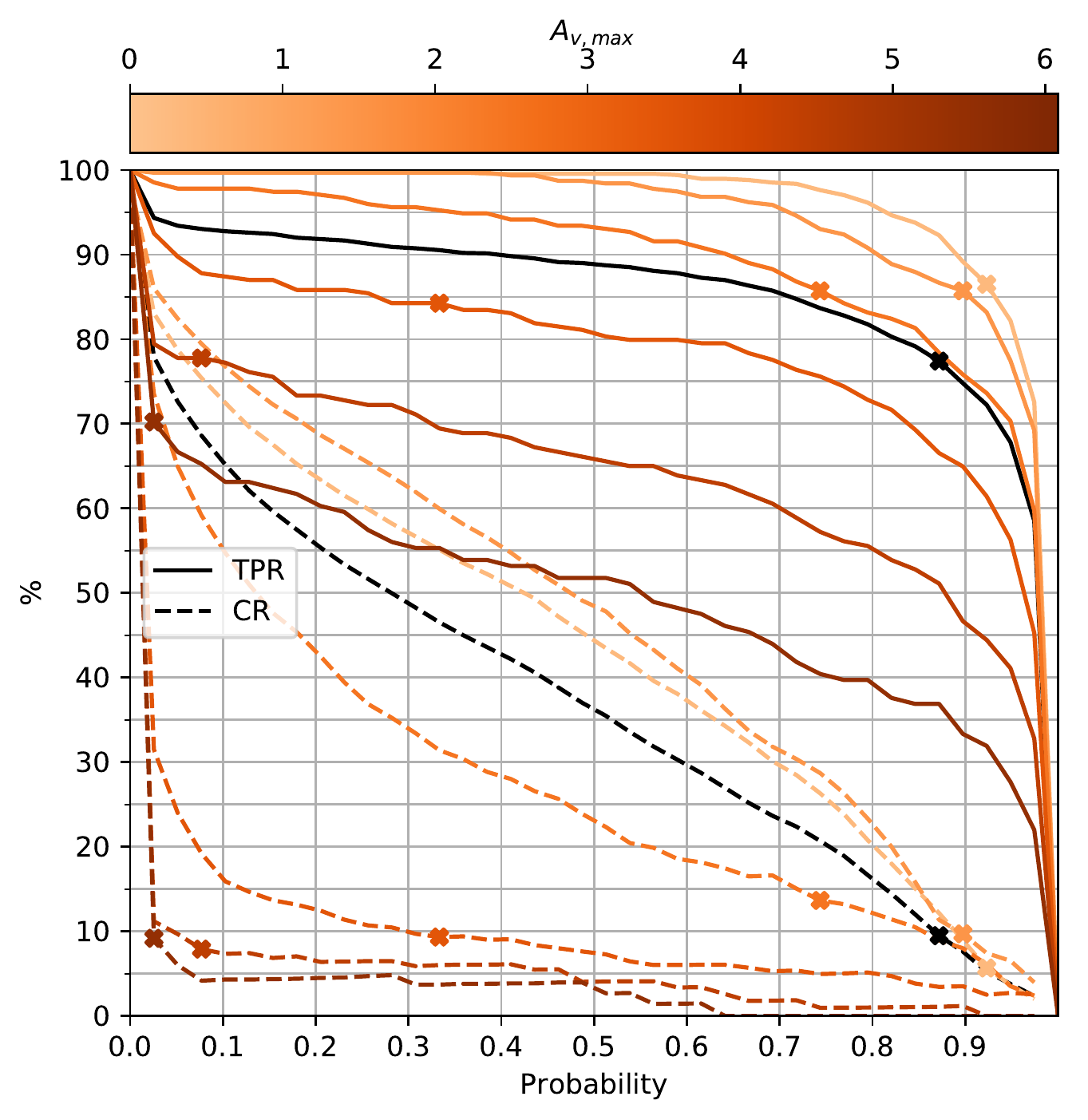}
     \caption{Quality indicators of the classifier obtained the MAP$_2$ data set. Captions as in Fig. \ref{fig:quality_map_0}.}
\label{fig:quality_map_2}
\end{figure}

Figures \ref{fig:quality_map_0} to \ref{fig:quality_map_2} show the TPR and CR of the classifiers obtained from the MAP$_{0,1,2}$ data sets as functions of the probability threshold. The black lines show the quality indicators obtained with the second strategy, that is using all sources, and the colored lines show those obtained with the third strategy and are color-coded with the mean extinction value of the bin. The mean extinction value and the optima probability thresholds of each case are shown in the second and third columns of Tables \ref{table:quality_map_0} to \ref{table:quality_map_2}. Similarly, the quality indicators at each optimum probability threshold are marked with crosses, and their values correspond to those shown in the fourth and seventh columns of the aforementioned tables. 

As can be observed from Fig. \ref{fig:quality_map_0}, the quality of the classifier resulting from the MAP$_0$ data set and the second strategy, shown with black crosses (TPR=93.9\% and CR=5.9\%), has similar quality indicators to those of the benchmark solution (i.e., TPR=92.5\% and CR=6.5\%), with both cases having accuracies $\gtrsim$99.6\%. The third strategy results in quality indicators for the first $A_v$ bin  that are similar to those of the second strategy. However, the second bin shows poorer results than the first one. Figures \ref{fig:quality_map_1} and \ref{fig:quality_map_2} show that the quality indicators found in the MAP$_{1,2}$ data sets follow similar trends, which are a decreasing TPR with increasing value of extinction, and a CR that remains $\lesssim$15\%. Moreover, the accuracy of our classifier over all data sets, extinction values, and strategies remain in the interval between 99.2\% and 99.8\%, thus proving its excellent quality.

We notice that a major caveat of our extinction methodology when used as a classifier is its difficulty in recovering members with increasing values of extinction. As shown in Figs. \ref{fig:quality_map_0} to \ref{fig:quality_map_2} and Tables \ref{table:quality_map_0} to \ref{table:quality_map_2}, the TPR steadily decreases as a function of extinction from values $\sim$90\% at $A_v\sim0.5$ mag, to $\sim$70\% at $A_v\sim5.5$ mag.  On the contrary, the CR resulting from the detailed analysis in bins of extinction (i.e., our third strategy) remains  $\lesssim$15\% in spite of the extinction value. 

\begin{figure}
    \centering
    \includegraphics[width=\columnwidth,page=2]{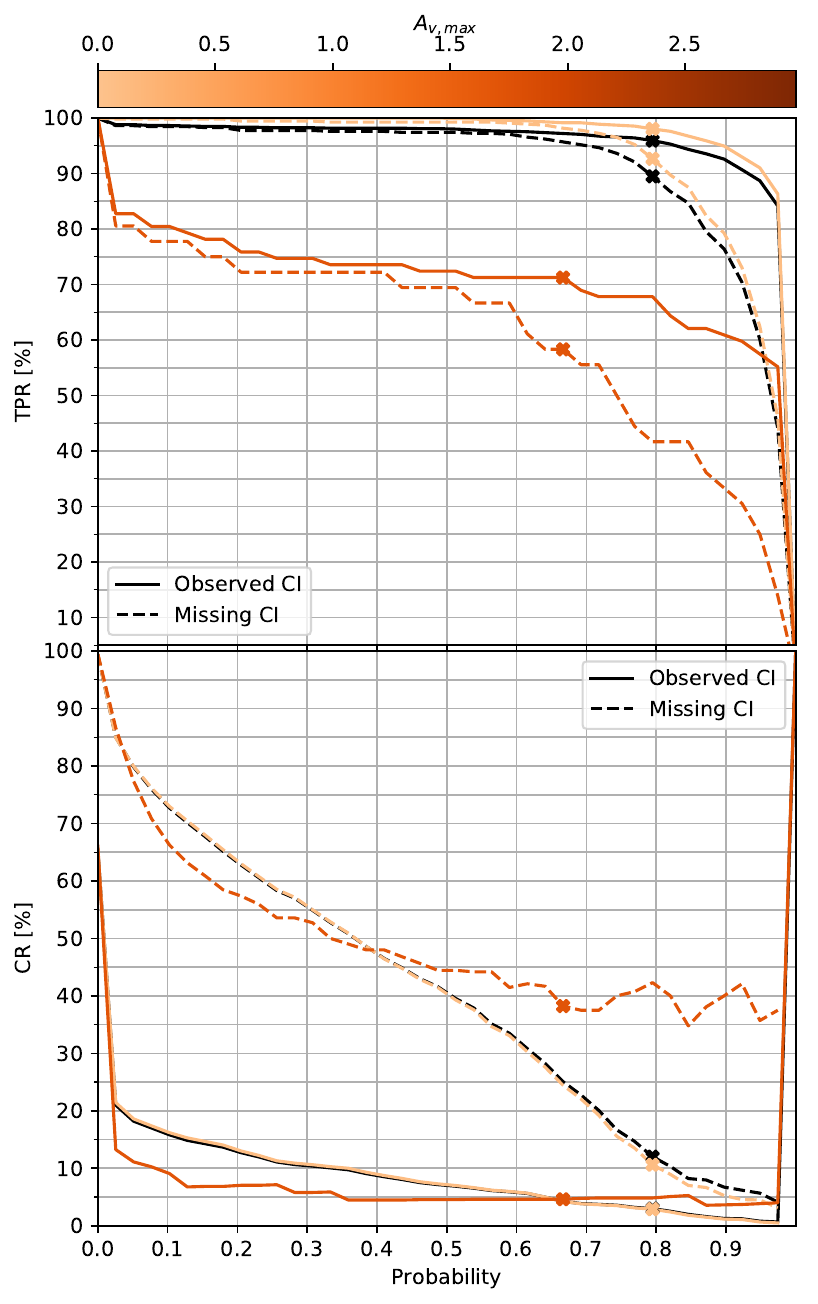}
     \caption{TPR and CR of the classifier resulting from the MAP$_0$ data set. The sources are split into those with a missing (dashed lines) and observed (solid lines) color index. The black line indicates the results independent of the extinction value of the sources (i.e., the second strategy), while the colored lines show the results of different bins of extinction (i.e., the third strategy).}
\label{fig:quality_nan_map_0}
\end{figure}

\begin{figure}
    \centering
    \includegraphics[width=\columnwidth,page=2]{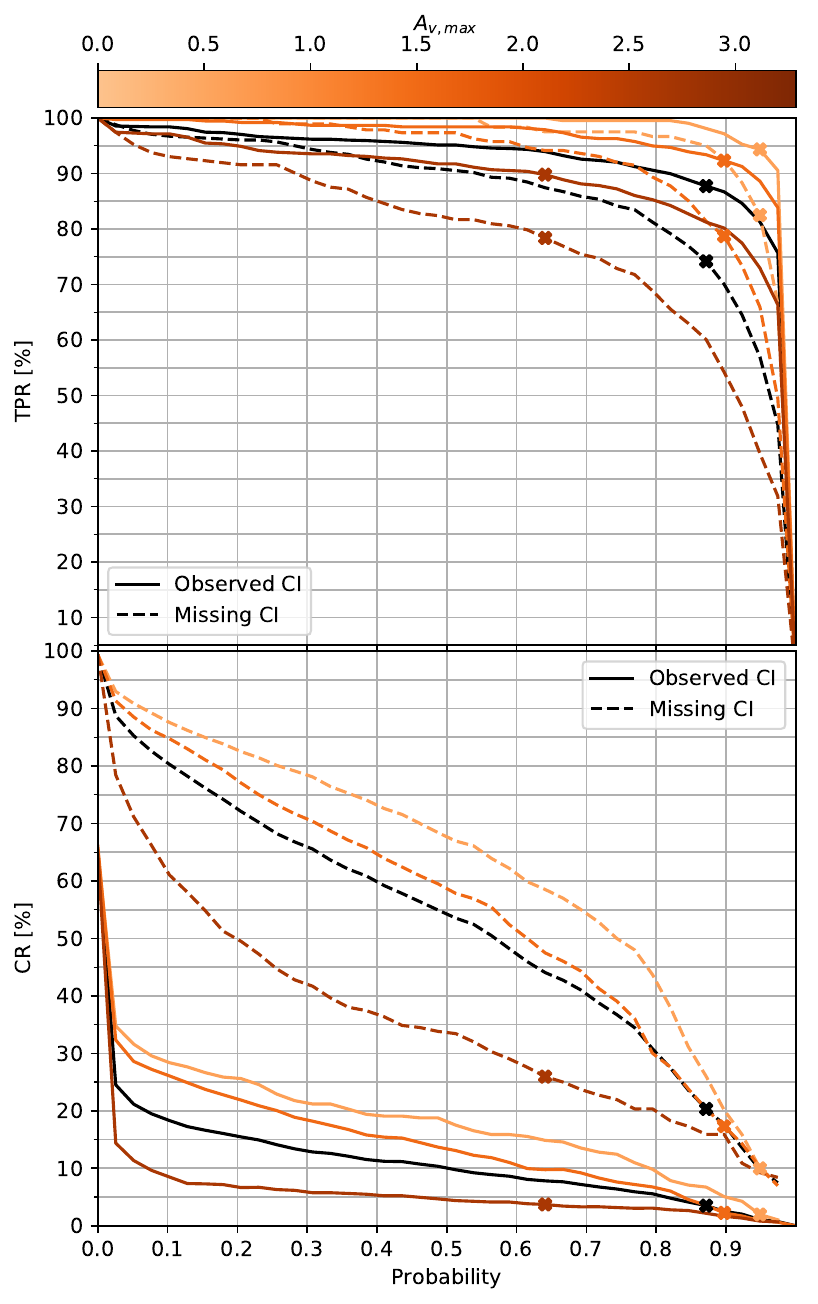}
     \caption{TPR and CR of the classifier resulting from the MAP$_1$ data set. Captions as in Fig. \ref{fig:quality_nan_map_0}.}
\label{fig:quality_nan_map_1}
\end{figure}

\begin{figure}
    \centering
    \includegraphics[width=\columnwidth,page=2]{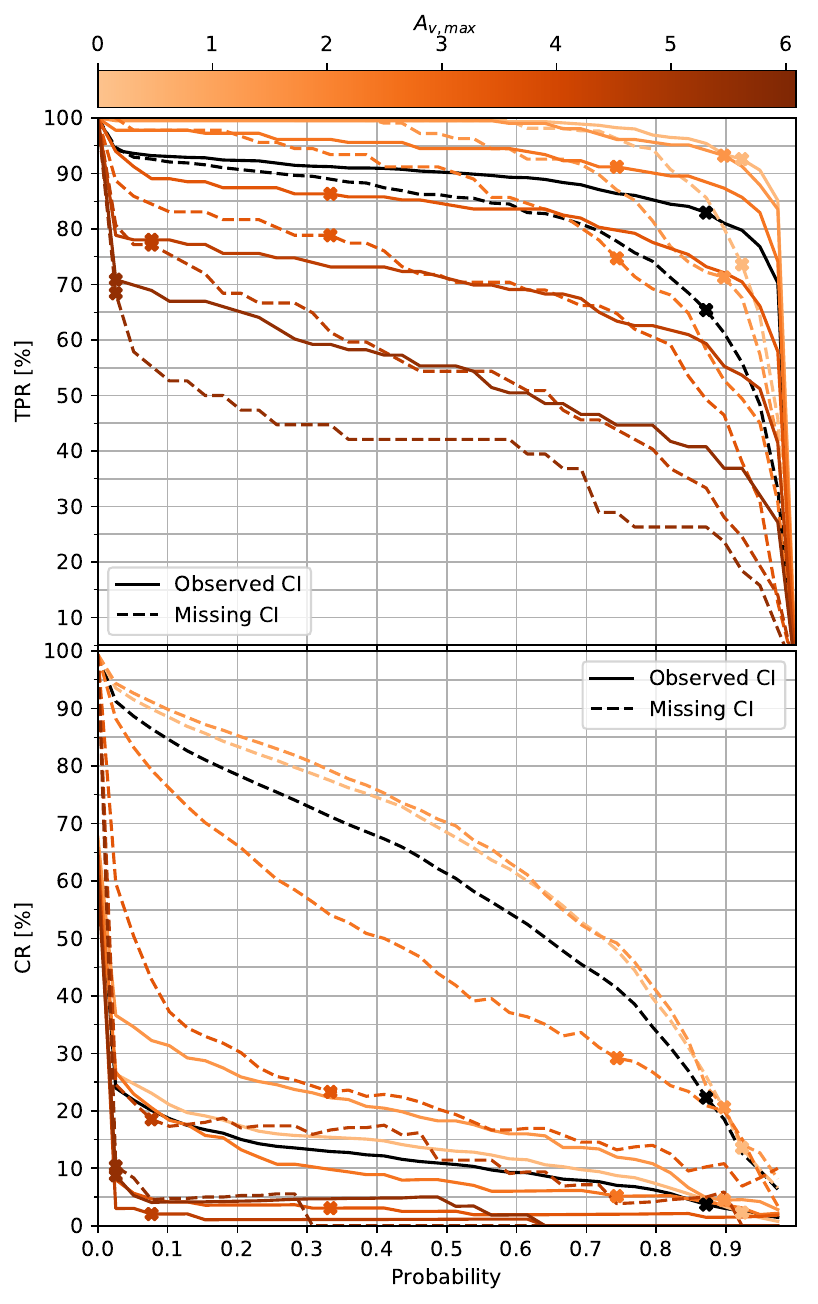}
     \caption{TPR and CR of the classifier resulting from the MAP$_2$ data set. Captions as in Fig. \ref{fig:quality_nan_map_0}.}
\label{fig:quality_nan_map_2}
\end{figure}

We now  analyze the quality indicators of the classifier for the subsets of the data with observed and missing color index. The results of this analysis are shown in Tables \ref{table:quality_map_0} to \ref{table:quality_map_2} and Figs. \ref{fig:quality_nan_map_0} to \ref{fig:quality_nan_map_2}. The upper and lower panels of these latter figures show the TPR and CR, respectively, as a function of the probability threshold, the extinction, and the status of the color index: observed or missing. 

As can be observed from Figs. \ref{fig:quality_nan_map_0} to \ref{fig:quality_nan_map_2}, the TPR of sources with observed color index is for all cases higher than that of sources with missing color index. The difference between these values decreases with increasing extinction, starting from 20\% at $A_v\sim0.5$, and down to zero at the limit of $A_v\sim6$, where the TPR of sources with and without a color index are both $\sim$70\%. Concerning the CR, we observe that sources with observed color index have consistently lower CR than those without it. In this case, the difference follows a trend similar to that of the TPR.

\begin{figure}
    \centering
    \includegraphics[width=\columnwidth,page=4]{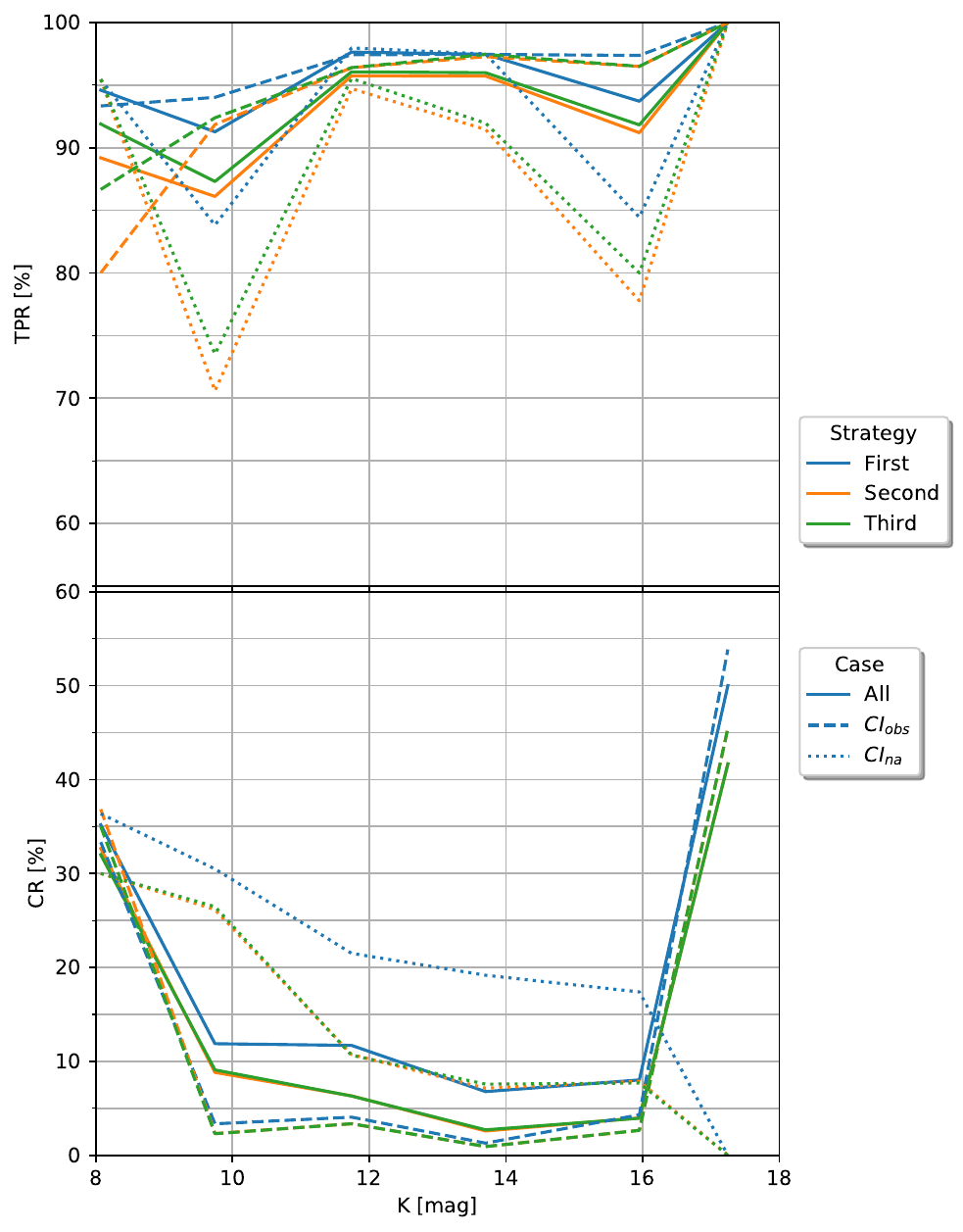}
     \caption{TPR and CR of the classifier resulting from the MAP$_0$ data set as functions of the K magnitude. The color shows the strategy to obtain the optimum probability threshold, and the line style shows the three different cases: all sources (solid lines), and those with and without an observed color index (dashed and dotted lines, respectively).}
\label{fig:quality_mag_map_0}
\end{figure}

\begin{figure}
    \centering
    \includegraphics[width=\columnwidth,page=4]{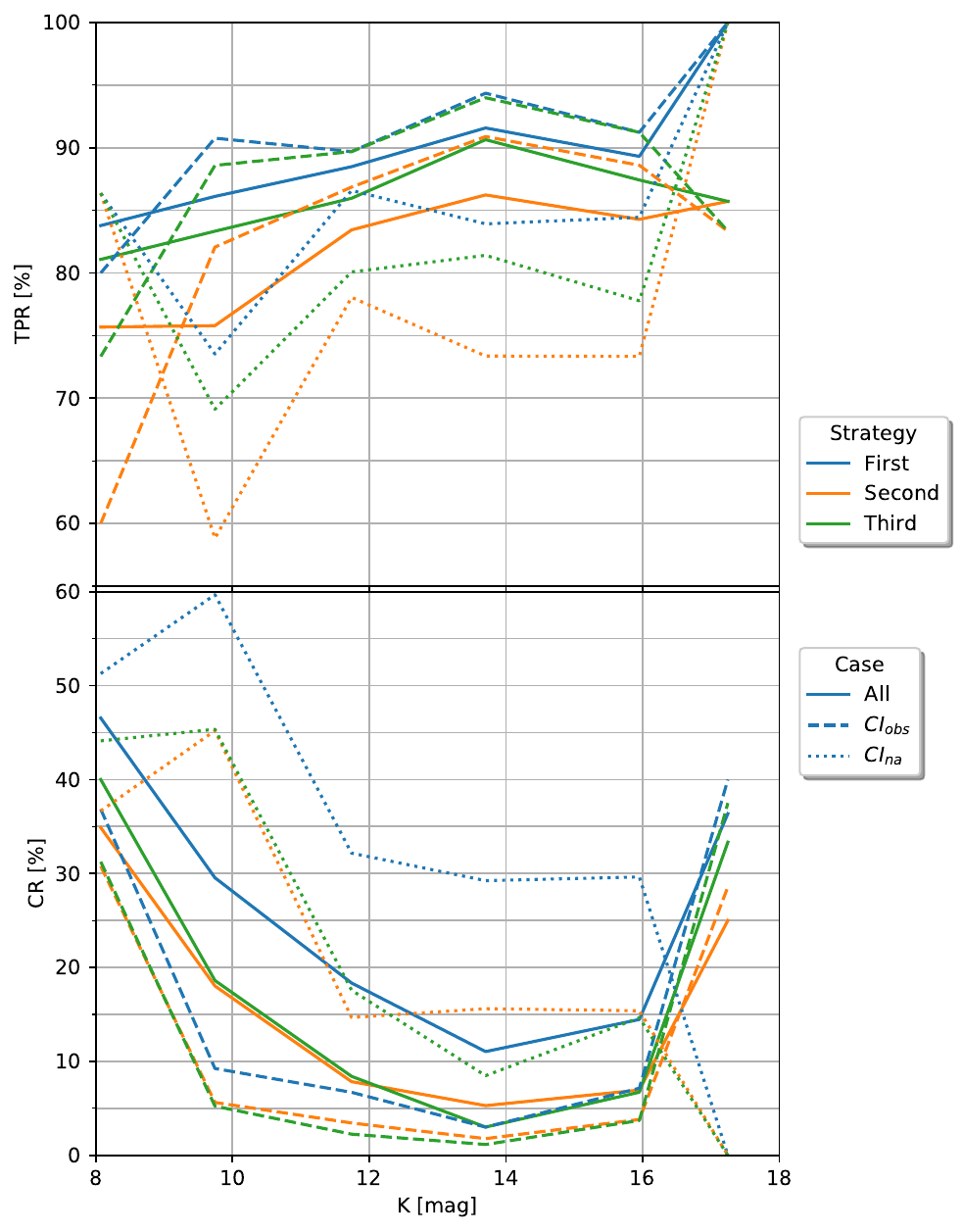}
     \caption{TPR and CR of the classifier resulting from the MAP$_1$ data set as functions of the K magnitude. Captions as in Fig. \ref{fig:quality_mag_map_0}.}
\label{fig:quality_mag_map_1}
\end{figure}

\begin{figure}
    \centering
    \includegraphics[width=\columnwidth,page=4]{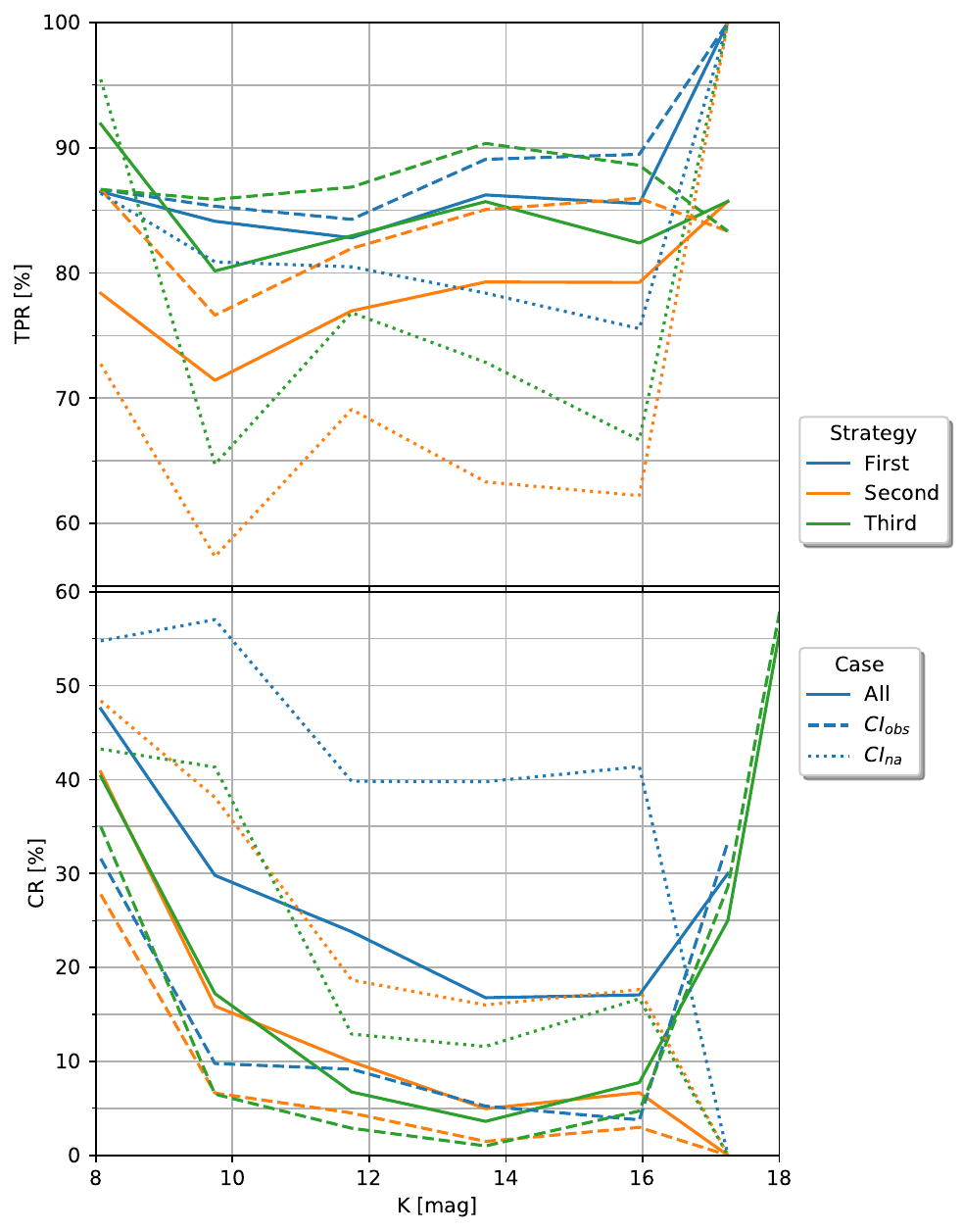}
     \caption{TPR and CR of the classifier resulting from the MAP$_2$ data set as functions of the K magnitude. Captions as in Fig. \ref{fig:quality_mag_map_0}.}
\label{fig:quality_mag_map_2}
\end{figure}

Finally, we analyze the dependencies of the TPR and CR on the photometric magnitude. Figures \ref{fig:quality_mag_map_0} to \ref{fig:quality_mag_map_2} show the measured TPR and CR of the classifier inferred from the MAP$_{0,1,2}$ data sets as functions of the K magnitude, which is the most observed band of our data set (see Table \ref{table:catalog}). In addition, they show the TPR and CR measured from the subsets of sources with and without an observed color index (marked as $CI_{obs}$ and $CI_{na}$, respectively). As can be observed in these figures, our three strategies show similar results of TPR, with the second one showing increasingly lower values of TPR with increasing extinction. Moreover, the TPR is relatively stable as a function of magnitude with slightly poorer values on the bright side (K$\lesssim$11 mag). The latter is a consequence of the high fraction of sources with missing Y band, an effect discussed in Sect. \ref{subsection:parameters} and Appendix \ref{appendix:classifier}. Regarding the CR, the values obtained with the second and third strategies are consistently lower than those obtained with the first one across all the magnitude interval and the data sets. In addition, the results from the three data sets show that the CR is lower in the central magnitude regions than in the faint and bright sides. However, we notice that due to low-number statistics, the results of the extreme bins at magnitudes <9 mag and >17 mag must be interpreted with caution.  As discussed in Sect. \ref{subsection:classifier}, sources with missing values hamper our methodology by increasing the CR. Therefore, the higher CR values on the bright and faint domains (without taking into account the edges) result from the large fraction of  sources with missing values produced by the sensitivity limits of the photometric detectors, which saturate at the bright side and are unable to detect the faintest sources. Partitioning the data set into the subsets with and without observed color index confirms our previous discussion about the negative impact that sources
with missing values produce in the classifier quality, particularly in the CR. As can be observed in these figures, the CR and TPR measured from the subset of sources with observed color index are better than those measured in the subset with missing color index. Here again, the values at the edges must be neglected due to the low-number statistics.

In this Appendix, we show that the major caveats of our methodology are the contaminants introduced by sources with missing color index, and the reduced recovery rate with increasing extinction value. These two problems are associated with the quality of both the data set and the extinction map. While better-quality data will certainly reduce the contaminants down to the negligible values shown by the subset with observed color index, a detailed extinction map that provides constraining lower and upper limits to the extinction will improve the TPR of the classifier. 

\end{appendix}
\end{document}